\documentclass[12pt,draftcls,onecolumn]{IEEEtran}
\usepackage{multirow}
\usepackage[utf8]{inputenc}
\usepackage{dblfloatfix}
\usepackage{bbm}
\usepackage{algpseudocode}
\usepackage{amsmath}
\usepackage{amssymb}
\usepackage{amsthm}
\usepackage{color}
\usepackage{cite}
\usepackage{graphicx}
\graphicspath{{./}{Figs/}}
\usepackage{subcaption}
\usepackage{mathrsfs}
\usepackage{verbatim}
\usepackage{indentfirst}
\usepackage{url}
\usepackage{epsfig,endnotes,amsthm,changepage}
\usepackage{authblk}
\usepackage[ruled,vlined]{algorithm2e}

\usepackage{epstopdf,lipsum,etoolbox}

\usepackage{graphicx}
\usepackage{amsfonts}
\usepackage{setspace}
\usepackage{cite}
\usepackage{array}
\usepackage{mdwmath}
\usepackage{mdwtab}
\usepackage{mdwtab, stmaryrd}
\usepackage{dsfont}
\usepackage{times}
\usepackage{epsfig}
\usepackage{latexsym}
\usepackage{epstopdf}
\usepackage{verbatim}
\usepackage{units}
\usepackage{amsthm}
\usepackage{mdwlist}
\usepackage{dblfloatfix}
\usepackage{blindtext}
\usepackage{array}
\newcolumntype{P}[1]{>{\centering\arraybackslash}p{#1}}
\usepackage{tabu}

\newcommand\Tstrut{\rule{0pt}{2.7ex}}

\AppendGraphicsExtensions{.pdf}
\usepackage{epstopdf,lipsum,etoolbox}

\newtheorem{definition}{Definition}

\newtheorem*{proposition*}{Proposition}

 \newcommand{\bF}{\mathbb{F}}

\allowdisplaybreaks

\begin{document}

\title{Private Retrieval, Computing and Learning: Recent Progress and Future Challenges \footnote{Sennur Ulukus is with University of Maryland (email: ulukus@umd.edu). Salman Avestimehr is with University of Southern California (email: avestimehr@ee.usc.edu). Michael Gastpar is with EPFL (email: michael.gastpar@epfl.ch). Syed Jafar is with University of California Irvine (email: syed@ece.uci.edu). Ravi Tandon is with University of Arizona (email: tandonr@email.arizona.edu). Chao Tian is with Texas A\&M University (email: chao.tian@tamu.edu).}}
\author{Sennur Ulukus, Salman Avestimehr, Michael Gastpar, Syed Jafar, \\ \vspace*{-0.4cm} Ravi Tandon, Chao Tian}

\maketitle

\vspace*{-1cm}

\begin{abstract}
Most of our lives are conducted in the cyberspace. The human notion of privacy translates into a cyber notion of privacy on many functions that take place in the cyberspace. This article focuses on three such functions: how to privately retrieve information from cyberspace (privacy in information retrieval), how to privately leverage large-scale distributed/parallel processing (privacy in distributed computing), and how to learn/train machine learning models from private data spread across multiple users (privacy in distributed (federated) learning). The article motivates each privacy setting, describes the problem formulation, summarizes breakthrough results in the history of each problem, and gives recent results and discusses some of the major ideas that emerged in each field. In addition, the cross-cutting techniques and interconnections between the three topics are discussed along with a set of open problems and challenges. 
\end{abstract}

\section{Introduction} \label{sect:intro}

Privacy is an important part of human life. This article considers privacy in the context of three distinct but related engineering applications, namely, privacy in retrieving information, privacy in computing functions, and privacy in learning. In the first sub-topic of private information retrieval, a user wishes to download a content from publicly accessible databases in such a way that the databases do not learn which particular content the user has downloaded. Towards that goal, the user creates ambiguity by its actions during the download. This strategy prevents databases from guessing which content the user has downloaded. This in turn preserves the user's privacy because what is downloaded leaks information about interest and intent on the part of the user. In the second sub-topic of private computing, a user wishes to compute a function but does not have resources to perform the computation on its own. Thus, the user outsources the computation to many distributed servers. This necessitates the user send its data, which is private, to the distributed servers. The goal of the user is to utilize the servers for computation while preserving the privacy of its own data. To achieve that goal, the user introduces randomness in its data so that the servers cannot decipher the data while they are able to perform the computation successfully. In the third sub-topic of privacy in learning, a centralized unit (parameter server) wishes to train a learning model by utilizing distributed users (clients). The parameter server needs labeled training data to train the model. The data resides at the users, and the users prefer to keep their data at their site, i.e., not send it to the parameter server, to preserve the privacy of their data. Thus, such distributed (federated) learning has built-in privacy advantages. However, even then, the computations (e.g., gradients calculated on the data) may leak some information about the raw data. To prevent that, the users may want to add randomness to the calculation they send to the parameter server in order to further preserve their privacy.      

The underlying threat model common to all three settings is the undesired leak of information that is considered private by the respective entities. In the case of private information retrieval, the leak is about the identity (index) of the content being downloaded/accessed. In the case of private computation, the leak is about the user data on which computation needs to performed by distributed servers. In the case of private learning, the leak is user (client) data that is used to train the learning model. A common aspect of the solution approach to these problems is to randomize the information/actions in such a way to hide the private information. In private information retrieval, this corresponds to randomizing the downloads such that a certain download may happen equally likely for all possible user content requirements. In private computation, randomization is achieved by adding appropriate noise to the data whose effect can be nullified during the computation. In private learning, privacy of clients is achieved by keeping the data at the client side, and also my randomizing the transmitted calculations so that leaks are prevented.  

We present private information retrieval in Section~\ref{sect:pir}, private distributed computation in Section~\ref{sect:pdc} and private distributed machine learning in Section~\ref{sect:pml}. We conclude this article in Section~Section~\ref{sect:conc} by listing a few challenges and open problems.

\section{Private Information Retrieval} \label{sect:pir}

The private information retrieval (PIR) problem was introduced by Chor et al. \cite{PIRfirst} as a privacy-preserving primitive for retrieving information in a private manner. In the canonical PIR setting, a user wishes to retrieve one of $K$ available messages, from $N$ non-communicating servers, each of which has a copy of these $K$ messages. User request privacy needs to be preserved during the retrieval process, i.e., the identity of the desired message remains unknown to any single server. A generic protocol to retrieve, e.g., message $k$, is as follows in this setting:
\begin{enumerate}
\item The user generates $N$ queries with a private random key and the message index $k$, one query per server, which are sent to the respective servers;
\item Each server, based on the query it receives and the content it stores, sends back an answer to the user;
\item The user collects the answers from $N$ databases, and reconstructs the message based on the answers, the private key, and the requested message index $k$.
\end{enumerate}

The privacy requirement can be either information-theoretic (e.g., \cite{Beimel_Ishai}) or computational (e.g., \cite{CPIR_log}). The former requires that each server cannot infer any information on the identity of the requested message, even if assuming the server has infinite computation power; in contrast, the latter assumes that each server has only limited computation power, and under such computational constraint, it is required that the server cannot learn anything on the identity of the requested message. In this article, we only consider information-theoretic privacy.

\begin{figure}[t]
\centering
\vspace{0.2cm}
\includegraphics[width=0.7\textwidth]{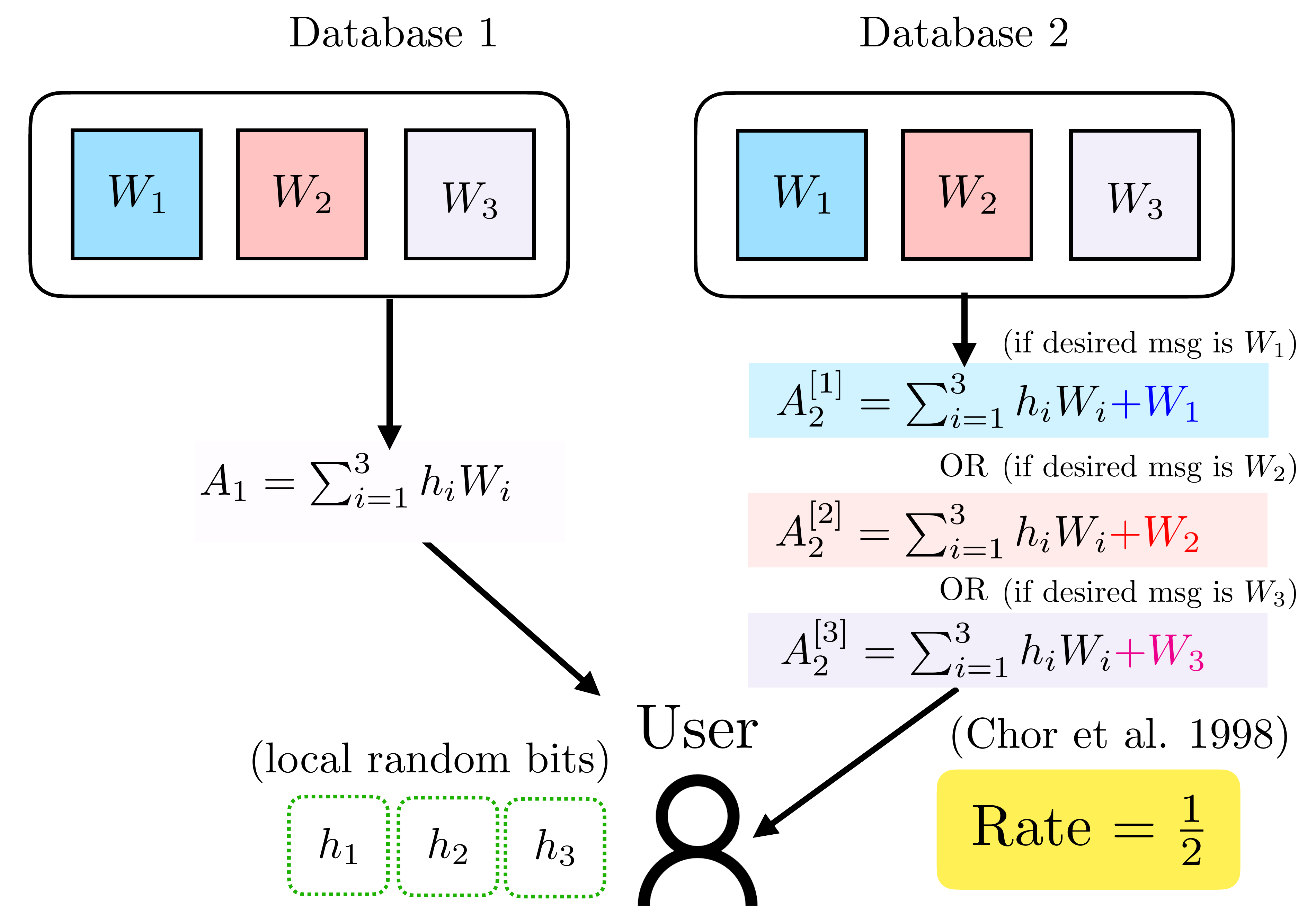}
\caption{The PIR scheme of Chor et al. \cite{PIRfirst} for $N=2$ databases which achieves a rate of $1/2$ for any number of messages $K$ (for the case shown in the figure, i.e., for $K=3$ messages, the capacity is $4/7$ \cite{sun2017PIRcapacity}). The main idea is to use the fact that since the databases cannot collude, one can send correlated queries across databases, allowing the user to leverage information retrieved across databases to increase the rate (download efficiency).  \label{fig:Chor-scheme} }
\end{figure}

Since the introduction of the PIR problem by Chor et al., tremendous advances have been made using the computer science theoretic approach, including on the canonical form and many variations; see the survey article \cite{William} and references therein. Within the context of this approach, the effort usually focuses on the scaling behavior of the communication costs, including both the query communication (upload) cost and the answer communication (download) cost, with respect to $N$ and $K$. Moreover, the messages are usually assumed to be very short, the most common of which is in fact a single bit per message. There have been many variations on the canonical setting, and it has been recognized that the PIR problem has deep connections to other coding or security primitives, such as locally decodable codes and oblivious transfer \cite{goldreich2002lower,YekhaninPhd,di2000single}. 

It was shown in Chor et al. that for a single database,  perfect information-theoretic privacy can only be achieved by downloading the entire database (of $K$ messages), i.e., the optimal rate defined as the ratio of the amount of desired information (one message)  and the total downloaded amount ($K$ messages) in this case is $1/K$. Thus, a natural question that was first explored by Chor et. al is the following: can one achieve a better rate than $1/K$ by exploiting $N>1$ databases? This question was answered in the affirmative, and it was shown that even with $N=2$ databases, one can achieve a rate of $1/2$ for any number of messages. We explain the main idea through a simple example as shown in Fig. \ref{fig:Chor-scheme} for $K=3$ messages. The main idea behind the scheme is as follows: assuming each message $W_k$ is one-bit, the user generates $K$ random bits $\{h_1, \ldots, h_K\}$
and  requests the linear combination (denoted by $\sum_{k=1}^{K}h_k W_k$) of the $K$ messages from database $1$, whereas requests $\sum_{k=1}^{K}h_k W_k + W_{\theta}$ from the other database whenever the user wants  to retrieve $W_\theta$. Since $\{h_k\}$'s are uniformly generated bits, the distribution of $\sum_{k=1}^{K}h_k W_k + W_{\theta}$ is identical for every $\theta \in \{1, 2,\ldots, K\}$, ensuring perfect privacy. 

\begin{figure}[t]
\centering
\vspace{0.2cm}
\includegraphics[width=0.7\textwidth]{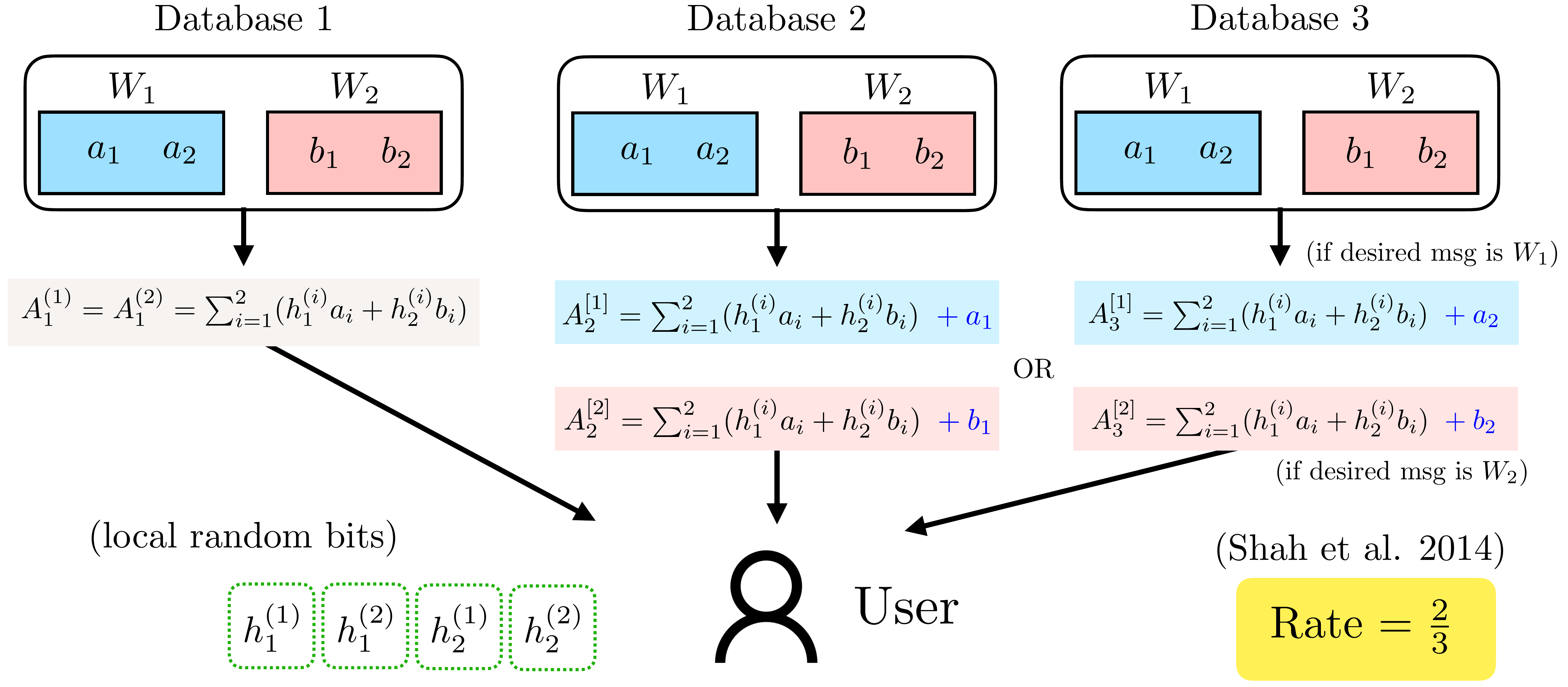}
\caption{The PIR scheme of Shah et al. \cite{Shah_Rashmi_Kannan} for $N=3$ databases and $K=2$ messages which achieves a rate of $2/3$ (the capacity for this setting is $3/4$ \cite{sun2017PIRcapacity}). More generally, the scheme achieves a rate of $1-1/N$, irrespective of $K$. The new ingredient beyond Chor et al. \cite{PIRfirst} involved message sub-packetization.  \label{fig:Shah-scheme} }
\end{figure}

The PIR problem was recently reintroduced to the information theory and coding community \cite{Shah_Rashmi_Kannan,Chan_Ho_Yamamoto,Fazeli_Vardy_Yaakobi}, with initial effort focused on using advanced coding technique to improve the storage, upload, and download efficiency. Specifically, Shah et al. \cite{Shah_Rashmi_Kannan} generalized the Chor et al. scheme to any arbitrary number $(N>1)$ of databases. The key new idea herein was to \text{subpacketize} each message into $(N-1)$ parts, and then follow an approach similar to Chor et al. In Fig.  
\ref{fig:Shah-scheme}, we highlight this through an example when $N=3$ and for $K=2$ messages. Each message is partitioned into $(N-1)=2$ parts, following which the user then requests a random linear combination ($A_{1}$ from database $1$), followed by requesting $A_{1}$ XORed with the $(N-1)$ subpackets individually from the remaining $(N-1)$ databases. This leads to a rate of $\frac{N-1}{N}=1-\frac{1}{N}$, which is independent of $K$. 

A significant milestone of this renewed effort on the PIR problem is the result obtained in \cite{sun2017PIRcapacity}, where the optimal download cost of the PIR capacity of the canonical setting was fully characterized. A key new ingredient that leads to this breakthrough is the information-theoretic reformulation of the problem. In contrast to typical computer science theoretical formulation, here the number of bits in each message is allowed to approach infinity, and the capacity is defined as the supremum of the number of useful message bits that can be retrieved per total downloaded bits. 

\begin{figure}[t]
\centering
\vspace{0.2cm}
\includegraphics[width=0.7\textwidth]{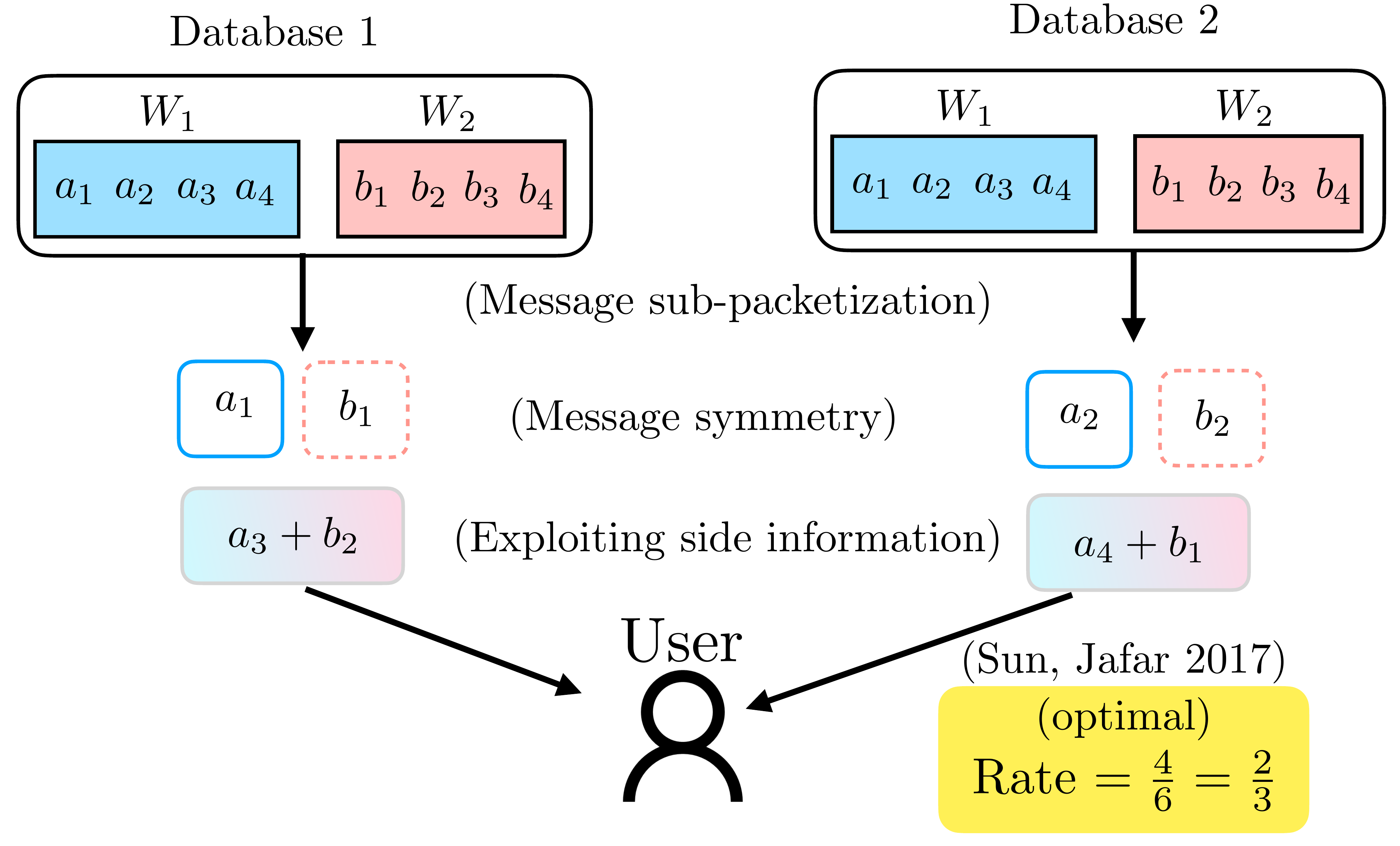}
\caption{The capacity-achieving PIR scheme of Sun and Jafar  \cite{sun2017PIRcapacity} for $N=2$ databases and $K=2$ messages which achieves a rate of $2/3$. The optimal scheme requires a combination of the following ideas: a) message sub-packetization, b) maintaining message symmetry for privacy and c) fully exploiting side information (from other databases) at each database.  \label{fig:Sun-Jafar-scheme} }
\end{figure}

A code construction was provided which relies on a few key code design principles. The scheme by Sun and Jafar for $N=2$ databases and $K=2$ messages is illustrated in Fig. \ref{fig:Sun-Jafar-scheme}. The key design principles behind the scheme include the following: a) sub-packetization of each message into $L= N^{K}$ symbols, followed by b) downloading parts of each message from every database (i.e., maintaining \textit{message symmetry}) for maintaining privacy of the desired message index, and c) fully exploiting side information at every database (from remaining $(N-1)$ databases). For the example in Fig. \ref{fig:Sun-Jafar-scheme}, this amounts to breaking the two messages into $L=4$ symbols. If the user wants to download message $W_1 = (a_1, a_2, a_3, a_4)$, it downloads one symbol from each message from both databases (namely, $(a_1, b_1)$ from database 1 and $(a_2, b_2)$ from database 2). Subsequently it downloads $a_3 + b_2$ from database 1 and $a_4+ b_1$ from database 2 (i.e., the remaining desired symbols together with the undesired symbols downloaded from the other database). A matching converse is proved using the conventional information theoretic approach. 

The surprising result inspired many subsequent works using such a capacity formulation and led to many new discoveries which will be surveyed in this part of the article. 

\subsection{The Canonical PIR System}

\begin{figure*}[t]
\centering
\vspace{0.2cm}
\includegraphics[width=0.98\textwidth]{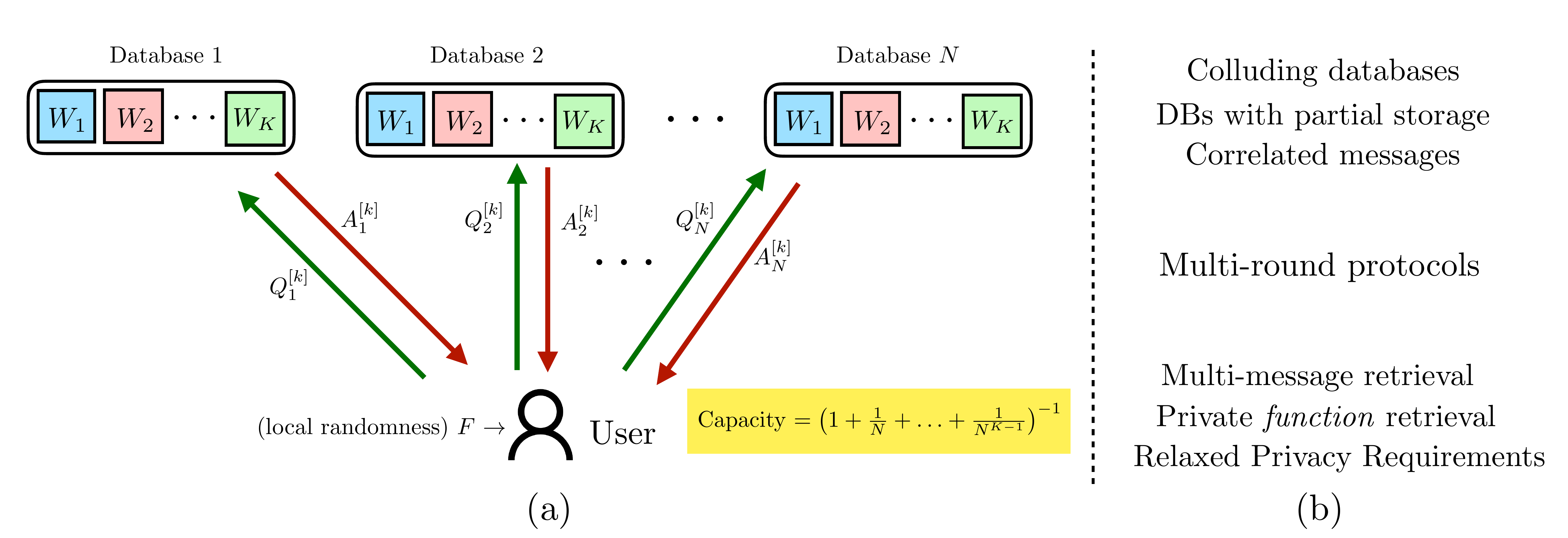}
\caption{(a) The canonical PIR system; (b) extensions of the canonical system. \label{fig:PIR_diag} }
\end{figure*}

For the canonical PIR setting, shown in Fig. \ref{fig:PIR_diag}, a rigorous computer science theoretic problem definition of the problem was given in the seminal paper \cite{PIRfirst}, and a more explicit information theoretic translation was given in \cite{tian2019capacity}. 
The breakthrough work of Sun and Jafar \cite{sun2017PIRcapacity} instead directly represented the coding function relations using information measure relations, which we explain next.

The random query $Q^{[k]}_n$ intended for server-$n$ when requesting message $k$ is determined by the private random key $\mathsf{F}$, i.e.,
$H(Q^{[k]}_n|\mathsf{F})=0$, for $n=1,2,\ldots,N,\, k=1,2,\ldots,K$. The answers $A^{[k]}_n$ from server-$n$, in response to the query $Q^{[k]}_n$, is determined by the stored messages and the query, i.e.,
$H(A^{[k]}_n|W_{1:K},Q^{[k]}_n)=0$,  for $n=1,2,\ldots,N,\,  k= 1,2,\ldots,K$. Given the above, there are two key constraints/requirements from a PIR scheme: 
\begin{enumerate}
\item \textit{Decodability Constraint:} The reconstructed message $\hat{W}_k$, by the user for the requested message $k$, is determined from the answers $A^{[k]}_{1:N}$ and the random key $\mathsf{F}$, i.e., 
\begin{align}
H(\hat{W}_k|A^{[k]}_{1:N},\mathsf{F})=0, \, \text{for } k\in\{1,2,\ldots,K\}.
\end{align}
\item \textit{Privacy Constraint:} The privacy requirement is that the queries for any message pairs $k$ and $k'$ have an identical distribution
\begin{align}
\mathbf{Pr}(Q^{[k]}_n=q)=\mathbf{Pr}(Q^{[k']}_n=q),
\end{align}
which can be represented as $I(\theta;Q^{[\theta]}_n,A^{[\theta]}_n,W_{1:K})=0$, for $n\in \{1, 2, \ldots, N\}$, where $\theta$ is the random variable representing the index of the requested message. 
\end{enumerate}

In the information-theoretic setting, the download cost $D$ dominates the upload cost. The definition of the download cost $D$ requires some elaboration. Two obvious information-theoretic measures directly related to the download cost are $\sum_{i=1}^N H(A^{[k]}_n)$ and $\sum_{i=1}^N H(A^{[k]}_n|\mathsf{F})$. The latter is a lower bound of the expected number of total download bits, which is how we usually measure the download cost. The former is an upper bound of the latter, and can be viewed as a surrogate, particularly in asymptotic (large number of information bits in each message) settings.  

The efficiency of the download is then measured by the number of requested message bits obtained per downloaded bit, which leads to the following capacity notion, when error is not allowed. More precisely, a rate $R$ is said to be achievable for zero-error PIR, if there exists a PIR code of download cost $D$ such that $R=\frac{L}{D}$ with no decoding errors. The supremum of achievable rates for zero-error PIR is called the (zero-error) PIR capacity $C_0$. 

For zero-error PIR code, there is no need to explicitly specify the probability distribution for each message, and also no need to specify the message retrieval probability. However, when a more general capacity notion, the $\epsilon$-error capacity, is adopted, this is no longer the case,  since the error probability is not strictly zero. In this case, the convention is to assume that each message is distributed in its range uniformly at random, and the message is also being requested uniformly at random \cite{sun2017PIRcapacity}. Correspondingly, a rate $R$ is said to be $\epsilon$-error achievable if there exists a sequence of PIR codes, each of rate greater than or equal to $R$, for which $\frac{1}{K}\sum_{k=1}^K\mathbf{Pr}(\hat{W}_k\neq W_k)\rightarrow 0$ as $L\rightarrow \infty$. The supremum of $\epsilon$-error achievable rates is called the $\epsilon$-error capacity $C_\epsilon$. 

The download cost used above is the expected number of downloaded bits, and the capacities are defined accordingly, which is usually the notion adopted in subsequent works. However another slightly different notion of the download cost is the worst-case download cost, which was used in \cite{sun2017optimal} (and later adopted in \cite{zhang2018optimal,jingke2017subScienceChina}). The worst-case download cost is the largest number of downloaded bits over all query combinations that are used with non-zero probability. Using this notion, we can similarly define the zero-error worst-case PIR capacity $\bar{C}_0$, and $\epsilon$-error worst-case PIR capacity $\bar{C}_\epsilon$. The breakthrough work \cite{sun2017PIRcapacity} essentially established that
\begin{align}
C_0=\bar{C}_0=C_\epsilon=\bar{C}_\epsilon=\left(1+\frac{1}{N}+\cdots+\frac{1}{N^{K-1}}\right)^{-1}.
\end{align}
The capacity definition $C_0$ is the most straightforward, and often adopted in the literature for generalized PIR settings. When the problem setting deviates from the canonical setting, it is known that $C_\epsilon$ can be different from $C_0$ in certain cases, but it is not well understood when this is the case. It is not known whether $\bar{C}_0$ and $\bar{C}_\epsilon$ can in fact be different for any generalized PIR systems.

\begin{figure}[t]
\centering
\vspace{0.2cm}
\includegraphics[width=0.9\textwidth]{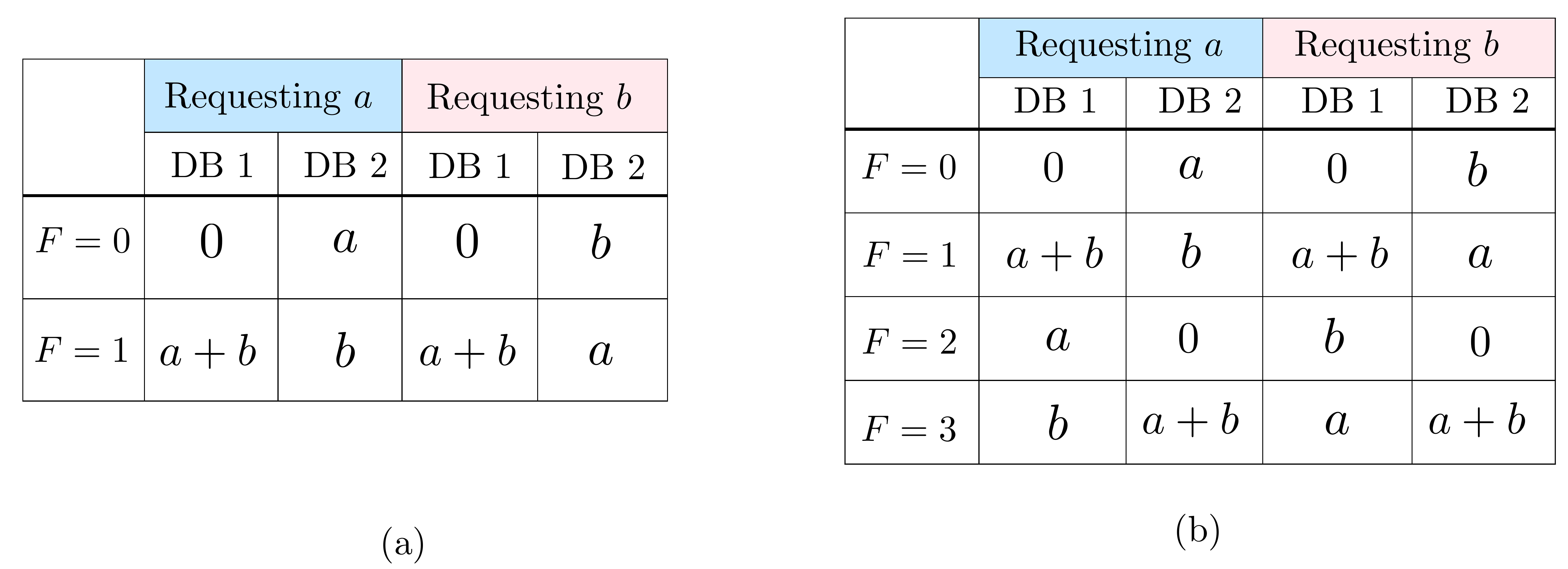}
\caption{Low-subpacketization schemes for PIR for $(N,K)=(2,2)$. The scheme in (a) was given in \cite{tian2018capacity_ICC,tian2019capacity}, and the scheme in (b) in \cite{Samy_tandon_lazos_leakyPIR_2019, asymmetric_leaky_PIR-2021}. One achieves the minimum possible subpacketization for each message while achieving an expected download cost of $3/2$.}
\label{fig:low-sub-pack-PIR}
\end{figure}

The general code construction for the canonical PIR system turns out to be rather elegant, and plays an important role for subsequent works. The code construction to obtain the capacity result in \cite{sun2017PIRcapacity} relies on several important design principles, which are illustrated using the example of $N=K=2$ case in Fig.~\ref{fig:Sun-Jafar-scheme}. This original construction required sub-packetization of each message into $N^{K}$ parts (alternatively, a message length of $L=N^{K}$ symbols), followed by invoking the design principles of server/message symmetry and exploiting side information. 

An alternative code construction was provided in \cite{tian2018capacity_ICC,tian2019capacity}, which is illustrated in Fig. \ref{fig:low-sub-pack-PIR}(a). In this construction, the server symmetry and and message symmetry are not used on the per retrieval basis, but across all the retrieval patterns. The random key $F\in\{0,1\}$ is invoked with probability $1/2$ each, and thus the expected retrieval download cost is still the same $3/2$. The advantage of this alternative code construction is that it can be shown to have the minimum message length ($L=1$ in this example and $L=N-1$ in general), comparing to the exponential growth of message length for the code given in \cite{sun2017PIRcapacity}. A similar code construction was discovered by \cite{Samy_tandon_lazos_leakyPIR_2019} for special case of $N=2$, and was later extended to the  case of more general number of databases \cite{asymmetric_leaky_PIR-2021} as shown in Fig. \ref{fig:low-sub-pack-PIR}(b). The difference from that in  \cite{tian2018capacity_ICC,tian2019capacity} is additional layer of symmetrization enforced across the databases.  It is clear that both code constructions have the same download cost, however, the upload cost of the former is lower ($\log 2$ vs $\log 4$). The symmetry structure in the canonical PIR setting is quite sophisticated and plays an important role in constructing efficient code design. The overall symmetry is induced by the database symmetry, the message symmetry, and the (retrieval) variety symmetry; see \cite{tian2019capacity} for a detailed discussion.

\subsection{Relation to Computer Science Theoretic PIR}

For the canonical PIR system, the computer science theoretic description of the coding operations is exactly equivalent to the information-theoretic version we just provided. The main difference between them is in terms of the performance measure, i.e., regarding the definition of $C_0$ and $C_{\epsilon}$. Since in the computer science theoretic setting the messages are short, usually only one bit each, the upload cost, i.e., $\sum_{i=1}^N\log |\mathcal{Q}_n|$, plays an important role in the overall communication cost, and thus the total communication cost must consist of both components. In contrast, in the information-theoretic setting, since the message size is allowed to be very large, the download cost dominates and the upload cost can be essentially ignored, and only the normalized cost is meaningful, whose inverse is the PIR rate. 

Since in the computer science theoretic setting, the message length is fixed and not allowed to grow to infinity, it is not meaningful to consider the ratio between the message length and the communication cost. In contrast, in the information-theoretic setting, this ratio between the message length and the download cost is the key metric to consider. 

It is in fact rather difficult to fully characterize the sum of the upload cost and the download cost (for any fixed message length), and thus the optimal scaling laws are usually sought after in the computer science theoretic setting. In contrast, in the information-theoretic setting, the ratio between the message length and the download cost leads to the concept of capacity, and the problem in fact becomes much more tractable. Instead of the scaling law, the capacity of the PIR system can be fully identified. 

\subsection{Extended PIR Systems} 

The canonical PIR system given in the previous sub-section can be viewed as consisting of four key components: a single-round query-answer protocol, a set of independent messages of the same length, an absolute privacy requirement, and also inherently a star-shape communication network; see Fig. \ref{fig:PIR_diag}. The last item regarding the communication network may require some elaboration, since it is usually not explicitly introduced: the user communicates to each server through a dedicated link, and each server answers through a dedicated link, and as a consequence, the communication costs are measured in a straightforward manner on each link. 

Any of these components can be generalized: 
\begin{enumerate}
\item The query-answer protocol structure. The user sends a single round of queries, and the servers answer in a single round; this protocol can be generalized to allow multiple rounds. 
\item The message structure.  In more general systems, the messages can be dependent; in a less obvious variation, the user in fact requests a function of the messages.
\item The privacy (or security) requirement.  The user may wish to enforce the privacy requirements that even certain subsets of the servers collude, they still will not be able to infer any knowledge on the request. The server may place security constraint that the authors cannot learn about other messages than the one being requested. 
\item The communication network structure.  This communication models can be generalized in various way, for example, to allow a more complex communication network, or using additional communication module, such as caches, to facilitate the communication. In such general settings, the communication costs are measured in rather different manners.
\end{enumerate}

In the next subsections, we  survey various generalizations  of the canonical PIR problem. 

\subsubsection{Multi-round and Multi-message PIR Systems}
Sun and Jafar  \cite{Sun_Jafar_MPIR} considered the extended PIR system where the user and the servers are allowed multiple rounds of queries and answers. It was shown that the capacity of multiround PIR is in fact the same as single round PIR, when there is no constraint placed on the storage cost. This equality continues to hold even when $T$-colluding is allowed. However, when the storage is more constrained, this equality would indeed break. Yao et al. \cite{yao2019capacity} considered using multiround communication in the settings with Byzantine databases, and showed that multiround communication is also beneficial in this setting. In multi-message PIR, the user wishes to download multiple messages privately. The question that arises is whether downloading multiple messages one-by-one sequentially is optimum. \cite{MMPIR} shows that downloading multiple messages jointly is more efficient and beats the sequential use of single-message PIR. \cite{MMPIR} determines the PIR capacity when the number of desired messages is at least half of the total number of messages, while the multi-message PIR capacity in other cases remains open. 

\subsubsection{Cache or Side Information Aided PIR}
Cache aided private information retrieval (PIR) (e.g., \cite{tandon2017capacity, PrefetchingPIR, Cache-aided_PIR}) and side information aided PIR (e.g., \cite{heidarzadeh2018capacity, KadheRouayheb2, ZhenWangJafar, PartialPSI_PIR, StorageConstrainedPIR_Wei, shariatpanahi2018multi, HeidarzadehRouayheb, li2018single, li2020singlemult, HeidarzadehSprintson}) are both interesting extensions of the original information-theoretic PIR problem \cite{sun2017PIRcapacity} because they both lead to reduction in download costs due to the fact that, under both settings, the user possesses cache or side information, respectively. The PIR capacity and the corresponding PIR schemes with cache/side-information vary, depending on a) if the databases are aware or unaware of the side-information at the user; b) if the user wishes to only keep the message index private or both the message index and side-information private from the databases; and c) the type of side-information available at the user (e.g., subset of messages or fraction of some/all messages). Recently, the role of side information is investigated in the context of symmetric PIR (SPIR) where the side information is a subset of shared database common randomness; this work showed that with appropriate amount of user-side side information the capacity of SPIR can be increased to the capacity PIR, and single-database SPIR can be made possible \cite{SPIR_UserRandomness}.

\subsubsection{PIR from Databases with Limited Storage}
The assumption of fully replicated databases (all $N$ databases storing all $K$ messages) can be unrealistic in practice. However, the amount of redundancy across the databases has an impact on the capacity of PIR. Specifically, on one extreme for replicated databases, the capacity is the highest, whereas on the other extreme, if there is no redundant storage across databases, then the only feasible strategy is to download all $K$ messages. There have been several recent works which have explored the trade-off between the capacity and storage for PIR. The case when each message is encoded by a maximum distance separable (MDS) code and stored across the databases, referred to as the MDS-PIR code, was studied in \cite{PIR_coded, Tajeddine_Rouayheb} and the capacity was settled by Banawan and Ulukus \cite{PIR_coded}; also see recent results on MDS-PIR with minimum message size \cite{zhou2020capacity}. The problem of MDS-PIR with colluding databases turns out to be more challenging  and the capacity remains unknown for general parameters; see \cite{FREIJ_HOLLANTI, Sun_Jafar_MDSTPIR}. Several other variants have been studied including PIR from databases storing data using an arbitrary linear code \cite{kumar2019achieving, Lin_Kumar_Rosnes_Amat}, impact on capacity versus storage when using arbitrary (possibly, non-linear codes) \cite{guo_ruida_tian_storage_PIR_2021, efficient_storage_ITW2019, tian_2019_storage_cost_journal,tian2018shannon,sun2019breaking}, when databases only store fraction of uncoded messages \cite{attia2018capacity, HeteroPIR, PIR_decentralized}, and when data is not perfectly replicated across the databases, but rather partially replicated according to graph based structures \cite{raviv2019GPIR, Karim_nonreplicated, Jia_Jafar_GXSTPIR,Sadeh_Gu_Tamo}.

\subsubsection{PIR Under Additional Abilities and Constraints for the Databases}
The original setting in \cite{sun2017PIRcapacity} considers privacy against individual databases. In practice, a subset of databases may have the ability to collude; this may happen, for instance, if the databases belong to the same entity. \cite{Sun_Jafar_TPIR} considers the case where up to $T$ out of $N$ databases may collude, and finds the PIR capacity as a function of $T$. Further, \cite{BPIRjournal} considers the case where in addition to the $T$ colluding databases, up to $B$ databases may exhibit Byzantine behavior, meaning that they can return arbitrarily random or incorrect answers to the queries, and finds the PIR capacity as a function of $T$ and $B$. In addition, databases may require \emph{database privacy}. This means that the user does not learn anything further than the message it wished to download. The resulting setting is coined as symmetric PIR (SPIR) to emphasize the symmetry of privacy requirements of the user and the databases. The capacity of SPIR is found in  \cite{Sun_Jafar_SPIR}. The SPIR capacity is smaller than the PIR capacity, as SPIR is a more constrained problem than PIR. SPIR achievable scheme is similar to the schemes in \cite{PIRfirst, Shah_Rashmi_Kannan} but it requires a shared common randomness among the databases. Recent paper \cite{SPIR_UserRandomness} explores making some of that common randomness available (randomly) to the user to increase SPIR capacity. Further, SPIR proves to be an important privacy primitive that is a building block in many problems that involve symmetric privacy requirements among participating parties, such as in private set intersection \cite{PSI_journal, MP-PSI_journal}. 

Databases may be subject to a set of practical limitations due to the way that the databases are accessed or the way they need to return their answers. For instance, if the databases need to return their answers via noisy and/or multiple-access wireless channels, then the PIR schemes should be designed together with channel coding techniques to deal with the uncertainty in the channels as in \cite{NoisyPIR}. In another example setting, if the rate at which the user can download information from the databases is different for each database, then the user access to the databases and the PIR schemes across the databases may need to be asymmetric. This may happen, for instance, if the databases have different distances to the user (with a more distant database having a smaller bit-rate) or if they have different channel qualities (some channels from the databases being in deep fades). In this case, asymmetric access conditions need to be taken into consideration \cite{AsymmetryHurtsPIR}. An interesting observation in \cite{AsymmetryHurtsPIR} is that if the asymmetry is mild, the full unconstrained PIR capacity may still be maintained. Another set of practical constraints arise if the database-to-user channels are being eavesdropped by an external entity. This gives rise to a problem formulation at the intersection of information-theoretic privacy and information-theoretic security \cite{SecurePIR, securePIRcapacity, PIR_WTC_II, securestoragePIR}. Yet another practical constraint is that the messages stored at the databases do not have to be of equal length, and their apriori probabilities of retrieval (popularities) do not have to be the same. These give rise to message semantics that need to be taken into account during a PIR code design \cite{SemanticPIR}. An interesting observation in \cite{SemanticPIR} is that if longer messages have higher popularities then the semantic PIR capacity may be larger than classical PIR capacity.

\subsubsection{Relaxed Privacy Notions}
Perfect information-theoretic privacy requirements (either for the user as in PIR or for both the user and the databases as in SPIR) usually come at the expense of high download cost and do not allow tuning the PIR efficiency and privacy according to the application requirements. In applications which may require frequently retrieving messages, trading user or database privacy for communication efficiency could be desirable. Ideally, one would select a desired leakage level and then design a leakage-constrained retrieval scheme that guarantees such privacy while maximizing the download efficiency. Asonov et al. introduced the concept of repudiative information retrieval  \cite{asonov2002repudiative}. The repudiation property is achieved if the probability that the desired message index was $i$ given the query is non-zero for every index $i$, i.e., there is always some remaining uncertainty at the database about the desired message index. Recently, Toledo et al. \cite{toledo2016lower} adopted a game-based differential privacy definition to increase the PIR capacity at the expense of bounded privacy loss. With the goal of allowing bounded leakage for the information-theoretic PIR/SPIR formulations (as initiated in \cite{PIRfirstjournal}), there have been a series of recent works. In \cite{Samy_tandon_lazos_leakyPIR_2019}, the perfect privacy constraint was relaxed by requiring that the log likelihood of the posterior distribution for any two message indices given the query is bounded by $\epsilon$. When $\epsilon=0$, this recovers perfect privacy, and allows leakage for $\epsilon>0$. Lin et al. \cite{lin2019weakly,lin2020capacity} relaxed user privacy by allowing bounded 
mutual information between the queries and the corresponding requested message index. Unlike \cite{lin2019weakly,lin2020capacity}, which deal with the average leakage measured by mutual information, the model studied in \cite{Samy_tandon_lazos_leakyPIR_2019} provides stronger privacy guarantees. Zhou et al. \cite{zhou2020weakly} measured the leakage using the maximal leakage metric and argued this leakage measure is more applicable. Guo et al. \cite{guo2019information} considered the problem of SPIR with perfect user privacy and relaxed database privacy. Database privacy was relaxed by allowing a bounded mutual information (no more than $\delta$) between the undesired messages, the queries, and the answers received by the user. Similar to the original work on SPIR in \cite{sun2018capacitysymmetric}, SPIR with relaxed database privacy in \cite{guo2019information} requires sharing common randomness among databases and comes at the expense of a loss in the PIR capacity. Asymmetric leaky PIR was explored in \cite{asymmetric_leaky_PIR-2021} where bounded leakage is allowed in both directions. Recently, the model of latent-variable PIR was introduced and studied, where instead of requiring privacy for the message index, one may require privacy of data correlated with the message \cite{latent-variable-PIR}. 

\begin{figure}[t!]
\begingroup
\fontsize{10}{12}\selectfont
\begin{tabular}{>{\small}p{0.9in}>{\small}p{2.3in}>{\small}p{2.5in}>{\small}p{0.1in}}\hline

\multicolumn{3}{c}{Full capacity characterization}\\\hline

\multirow{3}{1.4in}{PIR\cite{sun2017PIRcapacity}\\
Multiround\cite{Sun_Jafar_MPIR}\\
Computation\cite{Sun_Jafar_PC, Mirmohseni_Maddah}\\
}&\multirow{3}{2.5in}{\hspace{1cm}$C=\Psi(N,K)$}&\multirow{3}{3in}{
Multiround: allows sequential queries\\
Computation:  retrieves arbitrary linear\\ combinations of messages}&\multirow{3}{*}{\checkmark}\\\\\hline
TPIR\cite{Sun_Jafar_TPIR}&$C=\Psi((N-U)/T,K)$&$U$  { u}nresponsive servers&\checkmark\\  \hline
\multirow{3}{1in}{PIR\cite{Yao_Liu_Kang_P} with \\  arbitrary collusion pattern}&\multirow{3}{1in}{$C=\Psi(S^*, K)$ }&\multirow{3}{3in}{Arbitrary collusion pattern $P$,\\$S^*=\max_y 1_N^T y$, s.t. $B_P^Ty\leq 1_M, y\geq 0_N$\\ $B_P$: incidence matrix of  $P$}&\multirow{2}{*}{\checkmark}\\ \\ \hline 
\multirow{2}{0.9in}{Cache-aided PIR\cite{Tandon_CachePIR}}&\multirow{2}{*}{$C=\Psi(N, K)/(1-S/K)$}& \multirow{2}{3in}{PIR aided by local {\bf c}ache at user of size $S\times$ message size} &\multirow{2}{*}{\checkmark}\\\hline
\multirow{2}{0.9in}{PIR-SI\cite{Kadhe_Garcia_Heidarzadeh_Rouayheb_Sprintson, Li_Gastpar}}&\multirow{2}{*}{$C=\Psi(N, \lceil\frac{K}{M+1}\rceil)$}& \multirow{2}{3in}{{\bf S}ide {\bf I}nformation of $M$ messages at User,\\ Privacy of SI not required}&\multirow{2}{*}{\checkmark} \\\hline
\multirow{2}{0.9in}{PIR-PSI\cite{Kadhe_Garcia_Heidarzadeh_Rouayheb_Sprintson, Chen_Wang_Jafar, PartialPSI_PIR}}&\multirow{2}{*}{$C=\Psi(N/T, K-M)$}& \multirow{2}{3in}{{\bf S}ide {\bf I}nformation of $M$ messages at User,\\ $\theta$ and SI jointly $T$-private}&\multirow{2}{*}{\checkmark} \\\hline

\multirow{3}{0.9in}{SPIR\cite{Sun_Jafar_SPIR, Wang_Skoglund_SPIRAd}}&\multirow{3}{3in}{$C=(1-\frac{2B+\max(T,E)}{N})$\\$\qquad\qquad\cdot\mathbb{I}_{\left(\rho\geq \frac{2B+\max(T,E)}{N-2B-\max(T,E)}\right)}$}& \multirow{3}{3in}{ {\bf S}ymmetric security, \\$B$-{\bf B}yzantine servers, $E$-{\bf E}avesdroppers,\\
 Common randomness $\rho$ shared among servers}\Tstrut &\multirow{2}{*}{\checkmark}\\  \\\hline
 
\multirow{3}{0.9in}{Q-PIR, Q-STPIR\cite{Song_Hayashi_QTPIR}}&\multirow{3}{2.5in}{${C}=\min\left\{1, \frac{2(N-T)}{N}\right\}$ }&\multirow{3}{3in}{Quantum PIR, \\
allows symmetric security,\\ Servers  share an entangled state}\Tstrut&\multirow{3}{*}{\checkmark}\\[0.3cm]\\
\hline

\multirow{2}{1in}{B-TPIR\cite{BPIRjournal}}&\multirow{2}{2.5in}{$C=\left(1-\frac{2B}{N}\right)\Psi(\frac{T}{N-2B},K)$, $\footnotesize{N>2B+T}$\\ $C=\frac{1}{(2B+1)K}\mathbb{I}_{(N>2B)}$, $N\leq 2B+T$}&\multirow{2}{3in}{\hspace{0.5cm}$B$-{\bf B}yzantine servers}\Tstrut&\multirow{2}{*}{\checkmark}\\[0.1cm]
\hline
\multirow{1}{0.9in}{MDS-PIR\cite{PIR_coded}}&$C=\Psi(N/K_c, K)$& $(N,K_c)$ MDS Coded Storage\\\hline
\multirow{2}{1.5in}{PIR with limited \\ storage \cite{attia2018capacity, HeteroPIR, PIR_decentralized}}&\multirow{2}{*}{\hspace{1cm}$C=\Psi(\mu N, K)$, $\mu=\frac{t}{N}, t\in[N]$}& \multirow{2}{3in}{Each server stores no more than $\mu$ fraction of \\
database; heterogeneous sizes $\mu_i$; decentralized.} \\\hline
  \hline
\end{tabular}  

\begin{tabular}{>{\small}p{0.9in}>{\small}p{2.3in}>{\small}p{2.5in}>{\small}p{0.1in}}\hline
\multicolumn{3}{c}{Partial capacity characterization}\\\hline

\multirow{2}{0.9in}{Multimessage PIR\cite{MMPIR}}&\multirow{2}{2in}{${C}=\Psi(N,\frac{K}{M})$ if $K/M\in\mathbb{N}$\\ $C=\frac{MN}{MN+K-M}$ if $M\geq\frac{K}{2}$}&\multirow{2}{3in}{{\bf M}ulti-{\bf m}essage Retrieval\\ Retrieves $M$ out of $K$ messages}&\multirow{2}{*}{\checkmark}  \\[0.1cm]\hline

\multirow{2}{0.9in}{MDS-TPIR\cite{Sun_Jafar_MDSTPIR, FREIJ_HOLLANTI}}&\multirow{2}{2in}{${C}_l=1-(K_c+T-1)/N$\\ $C=\frac{N^2-N}{2N^2-3N+T}$ if $K_c=N-1$}& \multirow{2}{*}{$(N,K_c)$ MDS Coded Storage}\\ 
\\[0.2cm]\hline

\multirow{3}{0.9in}{XS-TPIR\cite{Jia_Sun_Jafar_XSTPIR}}&\multirow{3}{2.5in}{${C}_l=\left(1-\frac{X+T}{N}\right)^+=C_\infty$\\ ${C}^u=\left(1-\frac{X}{N}\right)\Psi(\frac{N-X}{T},K)$\\  $=C$ if {\footnotesize $N\leq X+T$} or {\footnotesize $(N,X,T)=(3,1,1)$}}&\multirow{3}{3in}{$X$-secure storage }\Tstrut\\[0.2cm]
 \\\hline

\multirow{2}{0.9in}{U-B-XS-MDS-TPIR\cite{Tajeddine_Gnilke_Karpuk_Hollanti, Jia_Jafar_MDSXSTPIR}}&\multirow{2}{2.5in}{${C}_l=\left(1-\frac{K_c+X+T+2B-1}{N-U}\right)$\\ $=C_\infty$ if $K_c=1,X=0$}&\multirow{3}{3in}{$X$-secure,\\ $(N, K_c+X)$ MDS coded storage, \\ $U$-{\bf u}nresponsive, $B$-{\bf B}yzantine servers\\ }\Tstrut\\[0.3cm]
\\\hline
\end{tabular}

\endgroup
\caption{A sampling of  capacity results for various forms of PIR, where $N$ is the number of servers, $K$ is the number of messages, and $T$ is the privacy parameter with default value being $1$. For the rows with a check mark, the storage on server is simple message replication. 
$C$ represents capacity, $C_\infty$ is the asymptotic capacity for large number of messages $(K\rightarrow\infty)$, ${C}^u$ and ${C}_l$ are upper and lower bounds on $C$, respectively. $\Psi(A,B)\triangleq (1+1/A+1/A^2+\cdots+1/A^{(B-1)})^{-1}$.}\label{table:capacity}
\end{figure}

\subsection{Connections to Other Security Primitives}

PIR holds particular significance as a point of convergence of complementary perspectives. 
It is well known that PIR shares intimate connections to prominent problems in theoretical computer science and cryptography, communication and information theory, and coding and signal processing. PIR protocols are often used as essential ingredients of oblivious transfer \cite{SymPIR}, instance hiding \cite{Hide, Hide_one, Hide_multiple},  multiparty computation \cite{Local_random}, secret sharing schemes \cite{Shamir, Beimel_Ishai_Kushilevitz_Orlov} and locally decodable codes \cite{YekhaninPhd}. Through the  topics of locally decodable, recoverable, repairable and correctable codes \cite{Gopalan_Huang_Simitci_Yekhanin}, PIR connects to  distributed data storage repair \cite{Dimakis_survey}, index coding \cite{Birk_Kol_Trans} and the entire umbrella of  network coding \cite{Ahlswede_Cai_etal} in general. PIR schemes are essentially interference alignment schemes \cite{Jafar_FnT} as the downloads comprise a mix of desired messages with undesired messages (interference). Efficient retrieval requires the alignment of interference dimensions across the downloads from different servers while keeping desired signals resolvable. It is not surprising then that interference alignment has been used implicitly in  PIR and index coding long before its applications in wireless networks \cite{Jafar_FnT}. Various equivalence results have been established between PIR and blind interference alignment (BIA) \cite{Sun_Jafar_PIR, Sun_Jafar_BIAPIR}; BIA and topological interference management (TIM) \cite{Jafar_TIM}; TIM and index coding \cite{Jafar_TIM}; index coding and locally repairable codes \cite{Shanmugam_Dimakis_LRC,Mazumdar_LRC}; locally repairable and locally decodable codes \cite{Gopalan_Huang_Simitci_Yekhanin}; and between locally decodable codes and PIR \cite{YekhaninPhd}. Add to this the equivalence between index coding and network coding  \cite{Rouayheb_Sprintson_Georghiades, Effros_Rouayheb_Langberg}, storage capacity and index coding \cite{Mazumdar}, index coding and hat guessing \cite{Riis_Hat}, or the application of asymptotic interference alignment schemes originally developed for wireless interference networks \cite{Cadambe_Jafar_int} to distributed storage exact repair \cite{Cadambe_Jafar_Maleki_Ramchandran_Suh}, and it becomes evident that discoveries in PIR have the potential for a ripple effect in their impact on a number of related problems.

The remainder of this article explores in greater depth the topics of secure distributed computing and private federated learning.

\section{Private Distributed Computing} \label{sect:pdc}
\subsection{Formulation}

\begin{figure}[t]
\centering
\includegraphics[scale=0.4]{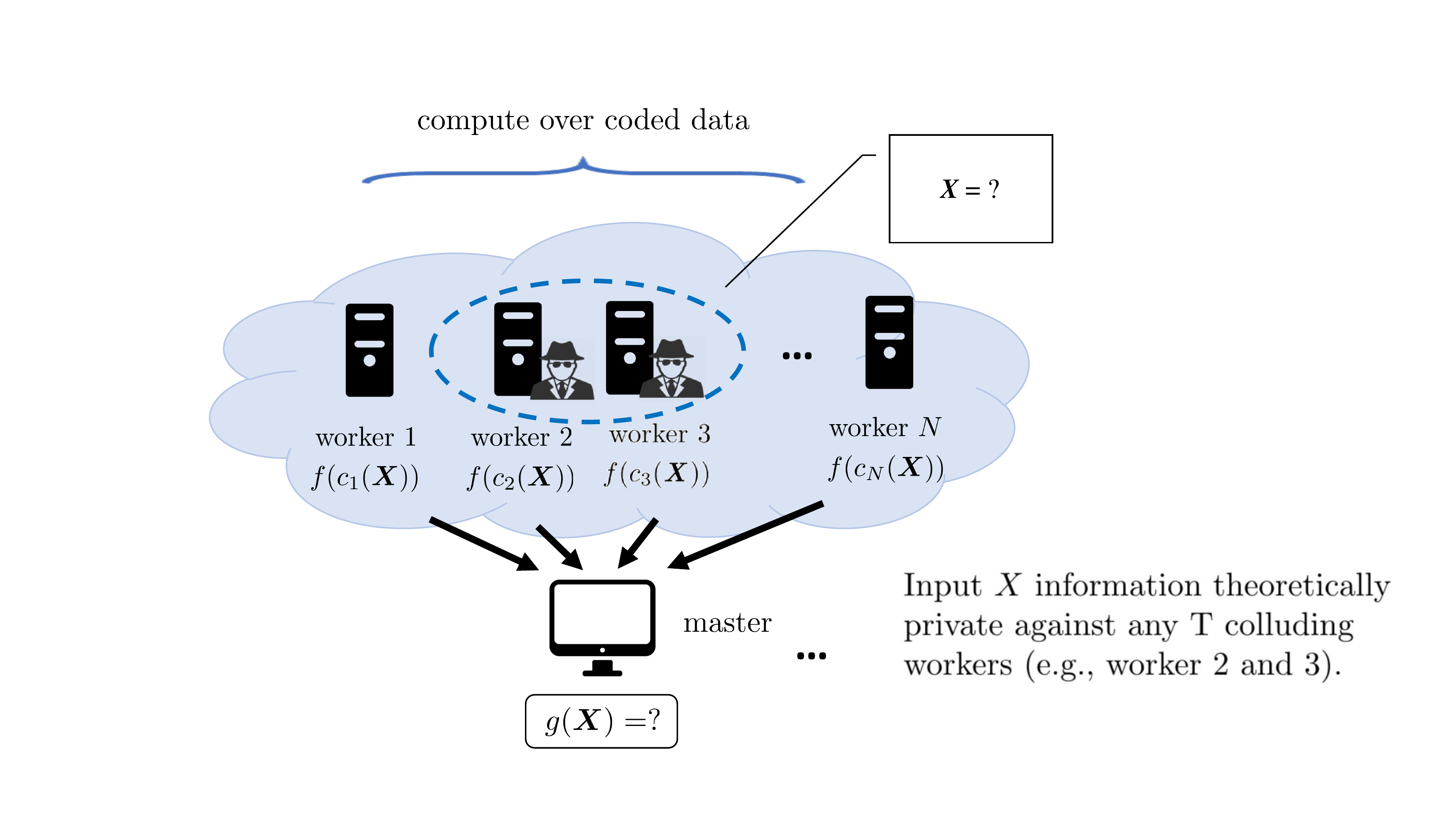}
\caption{An illustration of private computation. %A collection of workers aim to compute a function $g$ given an input dataset, where each worker can return an evaluation of a function $f$ with possibly coded data assignments. By carefully designing the coding functions ($c_i$'s), . 
}
\label{fig:overview}
\end{figure}

We consider a general distributed computing framework, where the goal is to compute a function $g$ using $N$ distributed workers, while keeping the input dataset $\boldsymbol{X}$ secure (illustrated in Fig.~\ref{fig:overview}). 
The $N$ workers are assigned encrypted versions of the input using $N$ encoding functions $\boldsymbol{c}\triangleq (c_1,...,c_N)$, 
then each worker computes a function $f$ over the assigned share, which can be viewed as building blocks of computing $g$.

This framework captures many commonly used operations.  One example is \emph{block matrix multiplication}, where the goal is to compute the product $A^\intercal B$ given two large matrices $A\in \bF^{s\times t}$ and $B\in \bF^{s\times r}$. 
Here the input dataset is $X=(A,B)$, and the computation task is $g(A,B)=A^\intercal B$. Given some partitioning parameters $p$, $m$, and $n$, the input matrices are partitioned block-wise into $p$-by-$m$ and $p$-by-$n$ sub-blocks of equal sizes, respectively. Then each worker is assigned a pair of sub-blocks and computes their product, i.e., the function $f$ is the multiplication of two matrices of sizes $\bF^{\frac{t}{m}\times \frac{s}{p}}$ and $\bF^{\frac{s}{p}\times \frac{r}{n}}$. If there are no security requirements, the final result can be recovered using $N=pmn$ workers, by having each worker compute a  product of certain uncoded submatrices.

Another example is to compute \emph{multivariate polynomials} on a dataset $X$. Particularly, given a general polynomial $f$, the input dataset is partitioned into $K$ subsets $X_1,\ldots,X_K$, and the goal is to compute $g(\boldsymbol{X})=(f(X_1),\ldots,f(X_K))$. If each worker can compute a single evaluation of $f$, then a computing design using $N=K$ workers can be obtained by assigning each worker a disjoint uncoded subset of the input.

However, in secure computing, we aim to carry out the computation with an additional requirement that the entire input dataset is information-theoretically secure from the workers, even if up to a certain number of them can collude. In particular, a set of encoding functions $\boldsymbol{c}\triangleq (c_1,\ldots,c_N)$ is $T$-secure, if 
    $I(\{c_i(X)\}_{i\in\mathcal{T}};X)=0$
for any subset $\mathcal{T}$ with a size of at most $T$, where $X$ is generated uniformly at random.

{\it Tradeoffs in secure distributed computing:} The goal is to design the encoding functions to achieve a tradeoff between the resources/constraints while ensuring the reliable recovery of the desired computation. Specifically, the resources could correspond to the number of available workers, storage and computation performed per worker. In addition to $T$-security, one may also be interested in communication efficient designs to account for bandwidth constraints (between master node and the workers). In the past few years, there have been significant progress in using ideas from coding/information theory to devise new schemes, and ultimately towards understanding these fundamental tradeoffs. As an example, a large body of work \cite{lee2015speeding, tandon2016gradient, yu2017optimally, dutta2016short, 8949560} have focused on the problem of distributed computing in the presence of stragglers. Here, since the overall latency of computation can be limited by the slowest workers, the goal is to design schemes which minimize the number of workers required to carry out the computation,  while satisfying the security requirement. 
More rigorously, let $\mathcal{C}_T$ denotes the set of allowable\footnote{The set $\mathcal{C}_T$ also captures practical constraints such as encoding complexities.} encoding function designs that are $T$-secure. Then, in a secure coded computing problem, given fixed parameters  $f,g,$ and $\mathcal{C}_T$, one aim could be to find computing schemes $\boldsymbol{c}\in \mathcal{C}_T$ that use as small number of workers $N$ as possible. Alternatively, when the total number of workers $N$ is fixed, one may be interested in the design of $T$-secure schemes with minimum download communication overhead. The capacity of secure distributed computation, analogous to that of PIR, can then be defined as the supremum of the ratio of the number of bits of desired information (the desired function, $g(\mathbf{X})$), to the total number of bits downloaded from the $N$ servers.

%In the presence of stragglers, the decoder waits for a subset of fastest workers until $g(\boldsymbol{X})$ can be recovered given the returned results from the workers using some decoding functions. 

%We say a coded computing scheme achieves a \emph{recovery threshold} of $R$, if the master can correctly decode the final output given the computing results from \emph{any} subset of at least $R$ workers. 
%This is an equivalent measure of the number of stragglers \new{(as well as the number of Byzantine adversaries)} that can be tolerated. 

%-
\subsection{Schemes for Private Distributed Computing}
Information-theoretically secure distributed computing has its origins in the celebrated work of Ben-Or Goldwasser Micali (BGW protocol) on tasks involving linear/bilinear computations. Specifically, the master node creates $N$ coded shares with $T$-secure guarantees (using Shamir's secret sharing scheme) which are subsequently sent to the workers. The workers subsequently compute the function on the coded shares. In a recent work \cite{Bitar-Secure-staircase}, \textit{staircase codes}, presented originally for a PIR problem \cite{bitar2018staircase}, were combined with the idea of secret sharing to minimize the overall latency for secure distributed matrix multiplication. 

Lagrange coded computing (LCC) \cite{pmlr-v89-yu19b} has been proposed to provide a unified solution for computing general multivariate polynomials. In comparison to the classical BGW (or similar Shamir's secret sharing based protocols), LCC reduces the amount of storage, communication and randomness overhead.   Given any fixed parameter $T$, LCC encodes the input variables using the following \textit{Lagrange interpolation polynomial}
\begin{align*}
    c(x)\triangleq \sum_{j\in[K]}X_j\cdot \prod_{k\in [K+T]\setminus\{j\}}\frac{x-x_k}{x_j-x_k}+
    \sum_{j=K+1}^{K+T} Z_j\cdot \prod_{k\in [K+T]\setminus\{j\}}\frac{x-x_k}{x_j-x_k},
\end{align*}
where $x_1,\ldots,x_{K+T}$ are some arbitrary distinct elements from the base field $\bF$, and $Z_i$'s are some random cryptographic keys generated uniformly\footnote{We assume that $\bF$ is finite so that the uniform distribution is well defined. } at random on the domain of $X_i$'s. Each worker $i$ selects a distinct variable $y_i$ from the base field that is not from $\{x_{1},\ldots,x_{K}\}$, and obtains $\tilde{X}_i\triangleq c(y_i)$ as the coded variable. LCC is $T$-secure, because the coded variables sent to any subset of $T$ workers are padded by an invertible linear transformation of $T$ random keys, which are jointly uniformly random. 
    
After each worker $i$ applies function $f$ over the coded inputs, they essentially evaluate the composed polynomial $f(c)$ at point $y_i$. On the other hand, the evaluations of the same polynomial at $x_1,\ldots,x_{K}$ are exactly the $K$ needed final results. Hence, by polynomial interpolation, the decoder can recover all final results by recovering $f(c)$, by receiving results from any subset of workers with a size greater than the degree of  $f(c)$. More precisely, let $\textup{deg} f$ denote the total degree of polynomial $f$, the degree of the composed polynomial equals $(K-1)\textup{deg} f$. Thus, LCC computes any multivariate polynomial with at most $N=(K-1)\textup{deg} f+1 $ workers. 

Secure coded computation has also been studied for different computation tasks and settings. A majority of works are on matrix multiplication \cite{chang2018capacity, 8382305, DBLP:journals/corr/abs-1810-13006, DBLP:journals/corr/abs-1812-09962, 8613446, DBLP:journals/corr/abs-1901-07705, kim2019private, 8675905 ,chang2019upload, 8761275, nodehi2019secure, jia2019capacity, kakar2019uplinkdownlink, aliasgari2019private, d2019degree, 9174167}, and it has been shown in \cite{8437563, 9174167} that for block-partition-based designs, the optimum number of workers to enable secure computation can be within a constant factor of a fundamental quantity called the bilinear complexity \cite{gs005}. Private gradient computation was studied in \cite{8849245} and it was shown that the optimal coding design is encoding the input variables using harmonic sequences. \cite{9333639,8849547} considered a setting where the workers send compressed versions of their computing results to tradeoff the download communication cost and the required number of workers. 

LCC has also been widely leveraged to enable privacy-preserving machine learning~\cite{so2021codedprivateml,copML}. In particular, authors in~\cite{so2021codedprivateml} have considered a scenario in which a data-owner (e.g., a hospital) wishes to train a logistic regression model by offloading the large volume of data (e.g., healthcare records) and computationally-intensive training tasks (e.g., gradient computations) to $N$ machines over a cloud platform, while ensuring that any collusions between $T$ out of $N$ workers do not leak information about the dataset. In this setting, CodedPrivateML~\cite{so2021codedprivateml} has been proposed, which leverages LCC, to provide three salient features:
\begin{enumerate}
\item it provides strong information-theoretic privacy guarantees for both the training dataset and model parameters. 
\item it enables fast training by distributing the training computation load effectively across several workers.
\item it secret shares the dataset and model parameters using coding and information theory principles, which significantly reduces the training time. 
\end{enumerate}
LCC has also been leveraged to break a fundamental ``quadratic barrier'' for secure model aggregation in federated learning~\cite{so2021turbo}. We defer the discussion on this topic to the next section. 

Within the scope of this article, the connection between PIR, secure distributed computing, and private federated learning is  exemplified by the idea of cross-subspace alignment (CSA) which extends to all three domains.  CSA codes originated in \cite{Jia_Sun_Jafar_XSTPIR} as a solution to XS-TPIR, i.e., the problem of $T$-private information retrieval from $N$ servers that store $K$ messages in an $X$-secure fashion. CSA codes then found applications in private secure coded computation \cite{Kim_Lee_PSCC, Chang_Tandon_PSDMM, Jia_Jafar_MDSXSTPIR}, and in particular secure distributed matrix multiplication (SDMM) \cite{Chang_Tandon_SDMMOS}. CSA codes were first applied to SDMM by Kakar et al. in \cite{Kakar_Ebadifar_Sezgin_CSA}, and subsequently applied to  secure distributed \emph{batch} matrix multiplication  (SDBMM) by Jia et al. in \cite{Jia_Jafar_SDBMM}. These works produced sharp capacity\footnote{Analogous to PIR, the capacity of SDBMM is defined as the supremum of the ratio of the number of bits of desired information (the desired matrix products), to the total number of bits downloaded  from the $N$ servers.} characterizations for various cases. For example, in \cite{Jia_Jafar_SDBMM} the capacity for $X$-secure distributed computation by $N$ servers of a batch of outer products of two vectors is shown to be $(1-X/N)^+$,  the capacity for computing the inner product of two  length-$K$ vectors is  $\frac{1}{K}(1-\frac{X}{N})^+$ when $N\leq 2X$, and for long vectors $(K\rightarrow\infty)$ the capacity of computing inner products is shown to be $(1-2X/N)^+$.  While LCC \cite{pmlr-v89-yu19b} codes and CSA codes originated in seemingly unrelated contexts of distributed secure computing and secure PIR, there are interesting connections between them. For example, in a special case of secure multiparty/distributed batch matrix multiplications, CSA codes yield LCC codes as a special case\cite{Jia_Jafar_CSA_MM, Chen_Jia_Wang_Jafar}. The generalization inherent in CSA codes is beneficial primarily in download-limited settings, where CSA codes are able to strictly outperform LCC codes. However, it is worth mentioning that to achieve order-optimal performances, entangled polynomial codes \cite{8437563, 9174167} should be applied, which enables coding over bilinear-complexity-based algebraic structures, and achieving order-wise improvements.  Finally, in the domain of federated learning, CSA codes were applied in \cite{Jia_Jafar_XSTPFSL} to find a solution to the problem of $X$-secure $T$-private federated submodel learning. Fundamentally, this is a problem of privately reading from and privately writing to a database comprising $K$ files (messages/submodels) that are stored across $N$ distributed servers in an $X$-secure fashion. The CSA read-write scheme of \cite{Jia_Jafar_XSTPFSL}  is able to fully update the storage at all $N$ servers after each write operation even if some of the servers (up to a specified threshold value) are inaccessible, and  achieves a synergistic gain from the joint design of private-read and private-write operations. Intuitively, the connection between these problems arises because the operation required at each server for many (but not all) PIR (and private write) schemes can be interpreted as a \emph{matrix multiplication} between a threshold-$T$ secret-shared query vector/matrix (polynomial encoded for $T$-privacy) and a threshold-$X$ secret-shared data vector/matrix (polynomial encoded for $X$-security), which produces various desired and undesired products. CSA codes are characterized by a Cauchy-Vandermonde structure that facilitates interference alignment of undesired products along the Vandermonde terms, while the desired products remain separable along the Cauchy terms. This alignment structure allows efficient downloads by reducing interference dimensions. Therefore, to the extent that a multiplication of polynomial encoded matrices is involved, and download efficiency is of concern, the same Cauchy-Vandermonde alignment structure facilitated by CSA codes turns out to be useful across these problems. It is also noteworthy that applications of CSA codes generalize naturally beyond matrix products, to tensor products, as seen in Double Blind Private Information Retrieval ($M$-way blind PIR in general) \cite{Lu_Jia_Jafar_DBTPIR}. 

\section{Private Federated Learning} \label{sect:pml}

\subsection{Threat Models for Private Federated Learning}
A typical federated learning (FL) system  \cite{pmlr-v54-mcmahan17a, AdvancesFLNOW2021} comprises users/workers, a server/curator, and an analyst, where users are connected to the server, and the server is subsequently connected to the analyst. Users wish to jointly train a machine learning model using their local datasets with the help of the server. The training is typically done using iterative algorithms such as gradient descent and its variants, where users receive the global learning model that needs to be trained from the server and compute gradients using their local datasets, and subsequently send the gradients or the updated local models back to the server for aggregation. 
The analyst may request for the model at any given time. Depending on who the malicious party is, its capability and intent, we can have several different threat models. For example, the analyst can be assumed to be honest but curious who does not actively attack the trustworthy server or users but tries to learn as much information about users as possible through the output released to it. Similarly, the server can also be assumed to be honest but curious. However, different from the analyst, the server can also be an active attacker, who alters the training process and/or baits users into revealing their information. Another possible threat model is when a subset of users is malicious, who try to tamper with the training process by sending altered gradients or model updates. We refer the reader to a recent excellent comprehensive survey on the subject of FL \cite{AdvancesFLNOW2021}, which gives an in-depth account of recent progress on various FL modalities, as well as challenges in achieving efficiency, privacy, fairness, and system level implementation. 

In this survey, we focus on the models where (a) the server is trustworthy and the analyst is honest but curious; and (b) both the server and the analyst are honest but curious. One may think that no information can be learned by the curious party due to the fact that the local data never leave the users, therefore, the local data is private. However, it has been shown that even gradients or updated models can be used to recover the data used during training for feed-forward neural networks \cite{Phong2017gradientLeak, Phong2018gradientLeak, Larochelle2020Inverting} and convolutional neural network \cite{Zhu2019DeepLeak, Wang2019Beyond}. This type of attack is known as gradient/model inversion attack. 

\begin{figure}[t]
    \centering
    \includegraphics[width=0.5\columnwidth]{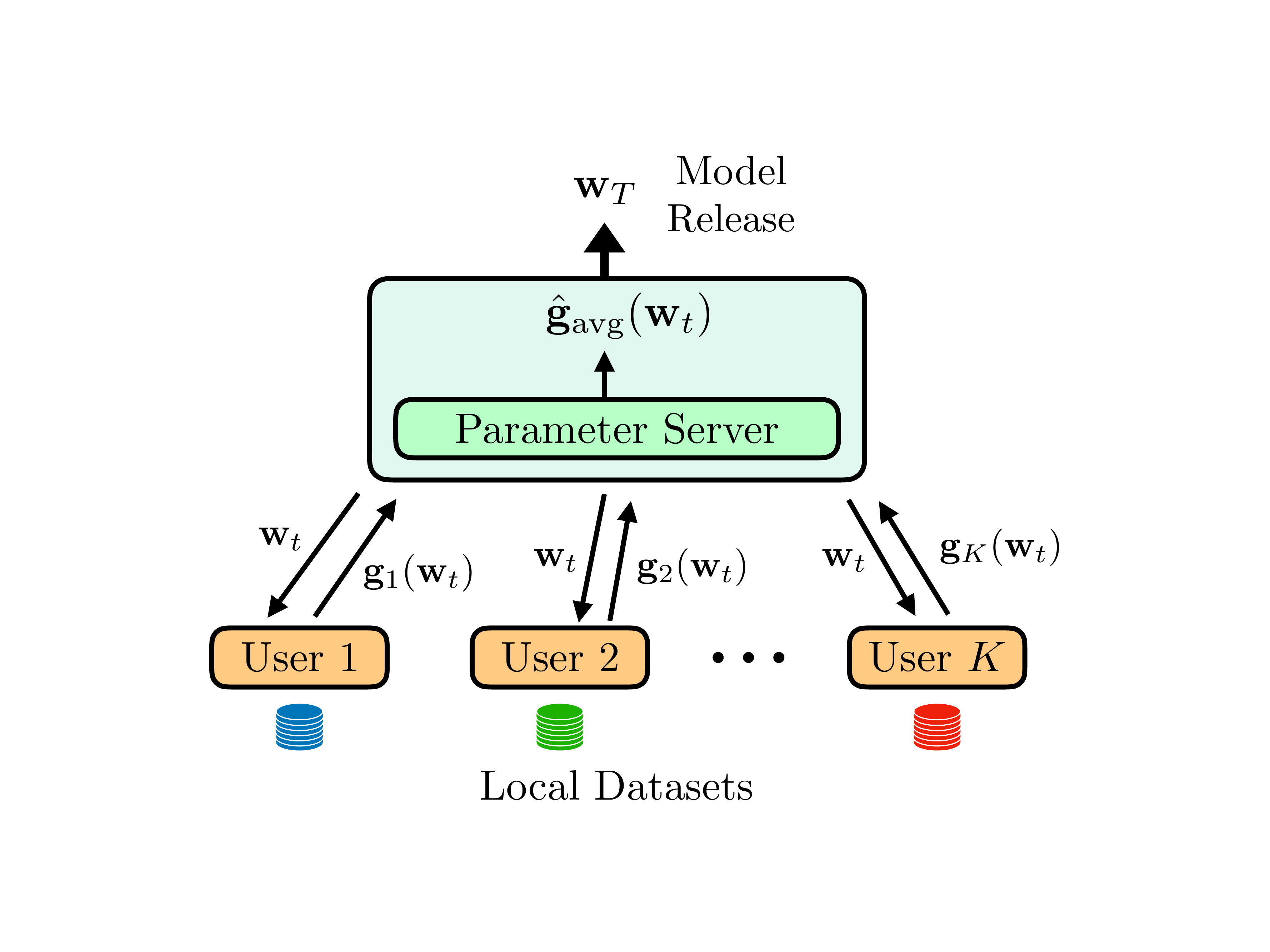}
    \caption{Conventional federated learning system, where $\mathbf{w}_t$ denotes the model parameters at iteration $t$, and $\mathbf{g}_k(\mathbf{w}_t)$ denotes the gradient computed using $\mathbf{w}_t$ by user $k$. After training, the model is released to an analyst.
    }
    \label{fig:FL_system_model}
\end{figure}

\subsection{Differential Private Federated Learning}

In private distributed computing, where the entire data is available at a central location (user), as discussed in the previous section, it is indeed possible to achieve perfect privacy (in an information-theoretic sense) when performing computations over distributed cluster of nodes. The federated learning paradigm, however, has several key distinctions as we briefly highlight next: since the data is already locally spread at the users (and is required to be kept private), perfect privacy against a single server can only be achieved by completely sacrificing utility (in terms of the model learned by perfectly private interactions with the user). Thus, in conventional single-server FL, one seeks to relax the privacy requirements from perfect privacy to allowing some leakage in a graceful manner. Indeed, as is shown in \cite{Jia_Jafar_XSTPFSL}, perfect privacy can be feasible with multiple servers, and when one may be interested in training multiple sub-models at the servers, or when some collaboration between the users is allowed (also see the discussion in Section~\ref{sec:securemodelaggregation}). For the remainder of this section, we will exclusively focus on the single-server FL setting when the users cannot collaborate.

Differential privacy (DP) \cite{DworkRothDPBook} is one of the most widely used privacy notions and has been shown to be effective to mitigate not only inversion attacks, but also differential attacks. The goal is to protect the private data by perturbing the output before it is released to untrustworthy parties. Depending on who performs the perturbation or who we wish to protect against, differential privacy can be further categorized into local DP and central DP. For a FL system with $K$ users, the local and central DP are formally defined as follows.

\begin{definition} 
($(\epsilon_{\ell}^{(k)}, \delta_{\ell})$-LDP) Let $\mathcal{X}_k$ be a set of all possible data points at user $k$. For user $k$, a randomized mechanism $\mathcal{M}_k: \mathcal{X}_{k} \rightarrow \mathds{R}^{d}$ is $(\epsilon_{\ell}^{(k)}, \delta_{\ell})$-LDP if for any $x,~x' \in \mathcal{X}_k$, and any measurable subset $\mathcal{O}_k \subseteq \text{Range}(\mathcal{M}_k)$, we have
\begin{align}
    \operatorname{Pr}(\mathcal{M}_k(x) \in \mathcal{O}_k) \leq \exp{(\epsilon_{\ell}^{(k)})}  \operatorname{Pr}(\mathcal{M}_k(x') \in \mathcal{O}_k) + \delta_{\ell}.
\end{align}
The setting when $\delta_{\ell} = 0$ is referred as pure $\epsilon_{\ell}^{(k)}$-LDP.
\end{definition}

\begin{definition} 
($(\epsilon_{c}, \delta_{c})$-DP) Let $\mathcal{D}\triangleq \mathcal{X}_1 \times\mathcal{X}_2\times\dots\times\mathcal{X}_K$ be the collection of all possible datasets of all $K$ users. A randomized mechanism $\mathcal{M}: \mathcal{D} \rightarrow \mathds{R}^{d}$ is $(\epsilon_{c}, \delta_{c})$-DP if for any two neighboring datasets $D, D'$ and any measurable subset $\mathcal{O} \subseteq \text{Range}(\mathcal{M})$, we have
\begin{align}
    \operatorname{Pr}(\mathcal{M}(D) \in \mathcal{O}) &\leq \exp{(\epsilon_{c})}  \operatorname{Pr}(\mathcal{M}(D')  \in \mathcal{O}) + \delta_{c}. 
\end{align}
The setting when $\delta_{c} = 0$ is referred as pure $\epsilon_{c}$-DP. 
\end{definition}
% Depending on the definition of neighboring datasets, central DP can be further categorized into example-level DP and user-level DP.

% with the worst privacy parameters, i.e., $\max_k \epsilon_{\ell}^{(k)}$. 
We refer $\epsilon_c$ ($\epsilon_{\ell}^{(k)}$) and $\delta_c$ ($\delta_{\ell}$) as privacy parameters. These parameters are closely associated with a quantity called sensitivity, which is defined as the largest difference of a function over all available inputs. It is known that central DP is a weaker guarantee than local DP. Therefore, local DP guarantee implies central DP guarantee. Both central and local DP are first computed on a per-iteration basis. Then, the total leakage is computed by summing up the leakages over all iterations. However, simply summing up the leakages over all iterations provides bound on the actual total leakage due to the fact that data is often reused during training. It is known that the more a data point is used, the more information it leaks. Therefore, to capture this phenomenon, various of composition theorem/leakage accountant methods are used to tighten the bound on the total leakage, such as advanced composition theorem, and moment accountant \cite{Abadi2016}. 

\begin{table}[t]
 \centering
  \begin{tabular}{|c|c|c|c|c|}
    \hline
     &  Noise Injection & Sampling & Shuffling & Others\\
     \hline
Central DP  & \cite{Agarwal2018cpSGD, Song2013SGD} & \cite{McMahan2018Recurrent, Asoodeh2020InfoTheory, Abadi2016, Bassily2014PrivateERM} & \cite{Erlingsson2019Shuffling, Bittau2017PROCHLO, Cheu2018Distributed} & \cite{Thakkar2019AdaptiveClipping} \\
Local DP  & \cite{Ono2020LDPReinforcementLearning, Li2020SecureFLwithDP} & \cite{Balle2018AmpSubsampling, Heikkila2020DPcrossSilo} & \cite{Erlingsson2019Shuffling, Balle2019Blanket, Ghazi2019Scalable, Balle2020RandomCheckIn, Girgis2021ShuffledModel} & \cite{Smith2017Interaction, Duchi2013Local} \\

         \hline
         
 \end{tabular}
%  \vspace{5pt}
 \caption{A quick reference of some of the key privacy preserving techniques for providing central and local DP guarantees in Federated Learning.
 \label{tbl:summary}}
\end{table}

\begin{enumerate}
    \item \textit{Basic privacy preserving mechanisms}: Let us first look at the case where the server is trustworthy, and the goal is to satisfy a desired central DP level against the curious analyst. The outputs, e.g., learning model iterates or gradients, have been shown to leak information about the local datasets. Therefore, the goal is to perturb the outputs so that it becomes difficult for the analyst to learn information about the local datasets. Typically, for works that focus on central DP, the perturbation is done at the server. The outputs can be perturbed by using random response, adding noise, or using approximations. For example, in \cite{Abadi2016}, Gaussian noise is added to the gradient before the model update. However, gradients and model parameters often are represented with finite precision, which make Gaussian noise injecting mechanism impractical. Thus, other types of noise injecting mechanisms are also considered, such as Laplace mechanism, and for discrete values, binomial mechanism can be used \cite{Agarwal2018cpSGD}. Noise with custom density can also be used \cite{Song2013SGD}. However, as mentioned earlier, the privacy guarantee degrades when the same data is used for training repeatedly. 
    
    \item \textit{Privacy amplification via sampling}: 
To remedy this issue, another line of work focuses on reducing the exposure of the data through user \cite{McMahan2018Recurrent, Asoodeh2020InfoTheory} or data point sampling \cite{Abadi2016, Bassily2014PrivateERM, Balle2018AmpSubsampling, Heikkila2020DPcrossSilo}. In \cite{McMahan2018Recurrent}, users are sampled i.i.d. according to some probability, who will then compute and send the gradients to the server for perturbation and model updates. In \cite{Bassily2014PrivateERM}, exponential mechanism is studied. In \cite{Balle2018AmpSubsampling}, various data sampling schemes, such as Poisson sampling, sampling with/without replacement, are studied and analyzed. As a result of sampling, the privacy level is \textit{amplified} \cite{Balle2018AmpSubsampling}, i.e., less noise is needed to achieve the same privacy level that is achieved by schemes without sampling. Amplification can also be obtained through shuffling \cite{Erlingsson2019Shuffling}, where a trusted shuffler shuffles the outputs from users before sending it to the server. Works that consider shuffling as part of the pipeline include \cite{Bittau2017PROCHLO, Cheu2018Distributed}. Another way to control leakage is to ensure that the sensitivity is small by carefully choosing the clipping norm \cite{Thakkar2019AdaptiveClipping}.

\begin{figure}[t]
    \centering
    \includegraphics[width=0.5\columnwidth]{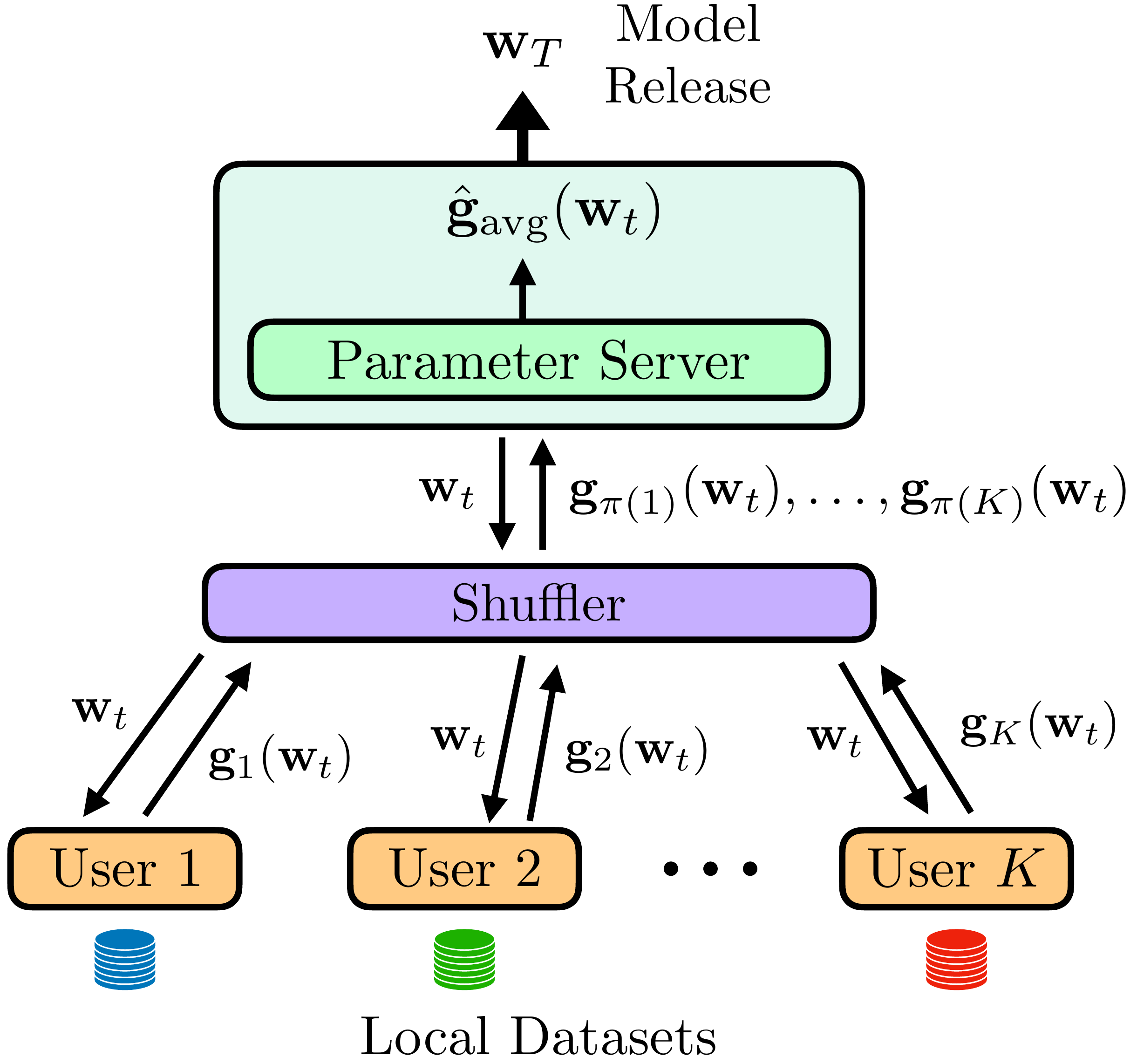}
    \caption{Federated learning system with a shuffler, who shuffles the gradients before sending them to the PS. The trained model is subsequently released to the analyst.
    }
    \label{fig:FL_system_model_shuffle}
\end{figure}

\item \textit{Private FL over new communication models}:
The above works rely on assumption that the server (and the shuffler) is trustworthy. This assumption may not be practical in certain scenarios. To remove this assumption, local DP was proposed and studied, where each user is responsible for protecting their own data. Similar to central DP, one can ask users to directly perturb the information they want to send, e.g., in \cite{Ono2020LDPReinforcementLearning}. In addition to adding noise, one can also perturb the information by using approximation. In \cite{Smith2017Interaction}, the approximation is obtained by flipping a random bit in the input string of each user. In \cite{Duchi2013Local}, a random vector that is roughly in the same direction as the original gradient is sampled and used as an approximation by each user. However, it has been shown that techniques that let users perturb information directly to achieve LDP may suffer greatly in terms of utility, i.e., accuracy. In order to satisfy LDP and provide reasonable utility, we can again use the idea of privacy amplification using sampling and/or shuffling \cite{Erlingsson2019Shuffling, Balle2019Blanket, Ghazi2019Scalable, Balle2020RandomCheckIn, Girgis2021ShuffledModel}. Since information is perturbed at the user, honest but curious shuffler would not compromise the local DP. Intuitively, both sampling and shuffling are able to further confuse the curious party without injecting more noise. Therefore, one is able to inject less noise to maintain utility, and still achieve the desired privacy level via sampling and/or shuffling. Another line of work focuses on private FL over wireless channels \cite{Seif2020Wireless, Sonee2020quantizedWirelessDP, Koda2020WirelessDP, Liu2021privacyForFree, Hasircioglu2021WirelessDP, Seif2021privacyAmp}. With superposition property of wireless channel, the gradients can be naturally aggregated while being transmitted. It has been shown in \cite{Koda2020WirelessDP} and \cite{Liu2021privacyForFree} that, by carefully designing the power control factors, the channel noise can be used as perturbation and provides DP guarantee. Amplification results are shown in \cite{Seif2020Wireless}, where perturbation added by users that is aggregated over wireless channel enhances privacy guarantee, and in \cite{Seif2021privacyAmp}, the privacy is amplified by aggregated perturbation and the addition of user sampling in the FL pipeline. However, channel state information (CSI) is obtained with the help of the server and is crucial in these works. When the server is untrustworthy, CSI obtained from the server can be tampered to lurk users to leak information. Therefore, \cite{Hasircioglu2021WirelessDP} and \cite{Seif2021privacyAmp} study the case when CSI is not available.

\item \textit{Other privacy notions and connections to DP}: There are also works that use different privacy notions, such as Concentrated DP \cite{Dwork2016CDP, Bun2016zCDP}, Renyi DP \cite{RDP}, Bayesian DP \cite{Triastcyn2019Bayesian}, communication-constrained DP \cite{Girgis2021ShuffledModel} and information-theoretic privacy \cite{so2021codedprivateml}. Concentrated DP is a relaxed version of DP, where it ensures that leakage is centered around the expected privacy level $\epsilon_c$, and is subgaussian. The probability that leakage exceeds $\epsilon_c$ by a small amount is bounded. Unlike the standard DP, where the expected leakage is not bounded and could potentially go to infinity with probability $\delta_c$, leakage of concentrated DP does not go to infinity. Renyi DP is another relaxation of the standard DP that is based on the concept of Renyi divergence. Renyi DP can be translated to standard DP and it was shown to have better composition result in \cite{RDP}. Bayesian DP is essentially standard DP, however, the data distribution is taken into account when quantifying privacy parameters. While one can show connection between DP and information-theoretic privacy, the approaches that are used to secure data are completely different. In \cite{so2021codedprivateml}, data is kept private by using error control codes and the idea of secret sharing. The data is considered private when the mutual information of the original data and encoded data is zero. Other works such as \cite{Guo2020DPDecentralizedLearning} studies how to allocate the amount of noise added to the data by each user in a decentralized setting (without the presence of a server) so that the collective noise does not reduce the utility.
\end{enumerate}

\subsection{Secure Model Aggregation in Federated Learning}\label{sec:securemodelaggregation}

While data is kept at the user-side in FL, a user's model still carries a significant amount of information about the local dataset of this user. Specifically, as shown recently, the private training data can be reconstructed from the local models through inference or inversion attacks (see e.g.,~\cite{fredrikson2015model,nasr2019comprehensive,zhu2020deep,geiping2020inverting}). To prevent such information leakage, \textit{secure aggregation} protocols are proposed (e.g., \cite{bonawitz2017practical,so2021turbo,kadhe2020fastsecagg,zhao2021information,bell2020secure}) to protect the privacy of individual local models, both from the server and other users, while still allowing the server to learn the aggregate model of the users. 
More specifically, secure aggregation protocols ensure that, at any given round, the server can only learn the aggregate model of the users, and beyond that no further information is revealed about the individual local model of a particular user. The key idea of the secure aggregation protocols is that the users mask their models before sending them to the server.  These masks then cancel out when the server aggregates the masked models, which allows the server to learn the aggregate of the local models without revealing the individual models.  

In the secure aggregation protocol of  \cite{bonawitz2017practical}, known as SecAgg, pairwise secret keys are generated between each pair of users. For handling user dropouts, the pairwise keys in \cite{bonawitz2017practical} are secret shared among all users, and can be reconstructed by the server in case of dropouts. This protocol tolerates any $D$ dropped users and ensures privacy guarantee against up to $T$ colluding users, provided that $T+D<N$, where $N$ is the number of users. The communication cost of constructing these masks, however, scales as $O(N^2)$, which limits the scalability of this approach. Several works have considered designing communication-efficient secure aggregation protocol \cite{so2021turbo,kadhe2020fastsecagg,zhao2021information,elkordy2020secure,bell2020secure}. SecAgg+ \cite{bell2020secure} improves upon SecAgg \cite{bonawitz2017practical} by limiting the secret sharing according to a sparse random graph instead of the complete graph considered in SecAgg  \cite{bonawitz2017practical}. TurboAgg \cite{so2021turbo} overcomes the quadratic aggregation overhead of \cite{bonawitz2017practical}, achieving a secure aggregation overhead of $O(N\log N)$, while simultaneously tolerating up to a user dropout rate of $50\%$ and providing privacy against up to $N/2$ colluding users with high probability. The key idea of Turbo-Aggregate that enables communication-efficient aggregation is that it employs a multi-group circular strategy in which the users are partitioned into groups. The dropout and the privacy guarantees of TurboAgg, however, are not worst-case guarantees and it requires $\log N$ rounds.  FastSecAgg \cite{kadhe2020fastsecagg} is a $3$-round secure aggregation interactive communication-efficient protocol that is based on the Fast Fourier Transform multi-secret sharing, but it provides lower dropout and privacy guarantees compared to SecAgg \cite{bonawitz2017practical}. While all of aforementioned works in secure aggregation provide cryptographic security, the secure aggregation protocol \cite{zhao2021information} provides information-theoretic security. In addition, unlike all previous protocols that depend on the pairwise random-seed reconstruction of the dropped users, this protocol departs from the previous protocols by employing instead one-shot aggregate-mask reconstruction of the surviving users. This feature can reduce the aggregation complexity significantly. However, this protocol relies on a trusted-third party to distribute the masks over the users.  While all of the aforementioned works do not consider the bandwidth heterogeneity among the different users in secure aggregation, an adaptive secure aggregation protocol has been proposed in \cite{elkordy2020secure} which quantizes the model of each user according to the available bandwidth to improve the training accuracy.
 
 While secure aggregation seeks to resolve the issue of preserving user data privacy by masking the individual model updates, the learning protocol can be adversarially affected by Byzantine users that may aim to break or perturb the learning to their benefit \cite{blanchard2017machine,so2020byzantine,he2020secure,prakash2020mitigating,khazbak2020mlguard}.  As the local models are protected by random masks, the server cannot observe the individual user updates in the clear, which prevents the server from utilizing outlier detection protocols to protect the model against Byzantine manipulations. This problem has been recently addressed in \cite{so2020byzantine}, for the I.I.D.~setting, where the first single-server Byzantine-resilient secure aggregation protocol for secure federated learning known as BREA has been developed. BREA is based on distance based adversarial detection and leverages quantization and verifiable secret sharing to provide robustness against malicious users, while preserving the privacy of the individual user models.

 All works on secure aggregation only guarantee the privacy of the individual users over a single aggregation round \cite{bonawitz2017practical,so2021turbo,kadhe2020fastsecagg,zhao2021information,bell2020secure}. While the privacy of the users is protected in each single round, the server can reconstruct an individual model from the aggregated models over multiple rounds of aggregation. Specifically, as a result of the client sampling strategy and the users dropouts, the server may be able to recover an individual model by exploiting the history of the aggregate models \cite{pejo2020quality, so2021securing}. This problem was studied for the first time in \cite{so2021securing} which  developed a client selection strategy known as Multi-RoundSecAgg that ensures the privacy of the individual users over all aggregation rounds while taking into account other important factors such as the aggregation fairness among the users and average number of users participating at each round (average aggregation cardinality) which control the convergence rate.

\section{Discussion: Challenges and Open Problems} \label{sect:conc}

In this article, we have surveyed the privacy issues in information retrieval, distributed computation, and distributed (federated) learning. We conclude this article with the following incomplete list of remaining challenges and open problems in these areas.

\emph{Challenges and open problems in private information retrieval:}

\begin{itemize}
\item Coded colluding databases: The PIR problem is completely solved when the database content is coded for the case when the databases do not collude, and is also completely solved when the databases collude for the case when the database content is replicated (uncoded). However, the problem is open when database content is coded, potentially secured and the databases may collude. Remarkably, even the asymptotic capacity (for large number of messages) remains open. While the lower bound for U-B-XS-MDS-TPIR in Table \ref{table:capacity} is conjectured to be aymptotically optimal, the asymptotic capacity $C_\infty$ remains unknown in almost all cases.
\item Non-replicated databases: The basic form of PIR assumes that the databases contain exactly the same set of files. In reality, the databases will have some overlap in content and will also have distinct items. When the databases have arbitrary contents, the PIR capacity problem is open, with a few notable exceptions\cite{raviv2019GPIR, Karim_nonreplicated, Jia_Jafar_GXSTPIR,Sadeh_Gu_Tamo}. The challenge here is to be able exploit the replication to reduce the download cost, while at the same time deal with non-replication as efficiently as possible. 
\item Upload cost and message size: In the capacity formulation, the upload cost is largely ignored. However, when the message sizes are not very large, the consideration on the upload cost becomes important. The problem then becomes how to construct codes for small message sizes to achieve a smaller upload cost. In general, PIR schemes should be designed to minimize a combined measure of upload and download costs. 
\item Weakly private (leaky) information retrieval: In some practical applications of PIR, it may not be absolutely necessary to require perfect privacy, and a small leakage may be tolerable. In this setting, two questions stand out: What are good metric(s) to measure the leakage, and how to characterize the capacity as a function of these metrics. 
\item More complex message structure: The messages are usually required to be independent (and of the same length). What is the optimal coding strategy when the messages are dependent, either as overlapping parts, or are dependent following a general probability law?
\item Privacy, stragglers and timeliness of retrieval: While PIR focuses mainly only on the privacy of downloaded information, it assumes that the servers are ideal, which respond to queries immediately and with no delays. Robust PIR problem considers the case of servers being completely unresponsive \cite{Sun_Jafar_TPIR, bitar2018staircase}. However, most servers respond eventually, albeit slowly in many cases. Therefore, there is a need to design systems where information may be downloaded privately but also in a timely manner. An initial consideration of this issue is presented in a recent paper in \cite{TimelyPIR}, but the problem remains largely open. 
\end{itemize}

\emph{Challenges and open problems in private distributed computing:}
\begin{itemize}
\item Optimal coding for block matrix multiplication: In a standard coded computing setup, we aim to find coding designs to minimize the recovery threshold \cite{NIPS2017_7027} (or the number of workers when no straggler is present) given fixed constraints on computation, security, and privacy. While the optimal recovery thresholds have been characterized within a factor of $2$ for block matrix multiplication  \cite{8437563, 9174167}, its exact characterization is not known except for some boundary cases. In particular, a remaining interesting open problem is to show whether the factor-of-$2$ penalty in the state-of-the-art upper bound is necessary when all partition parameters are large.
\item Analog coded computing: Most works in coded computing such as \cite{NIPS2017_7027,pmlr-v89-yu19b,9174167,soleymani2021list,8437563} rely on quantizing the data into finite fields and then leveraging coding-theoretic techniques to mitigate stragglers and Byzantine workers, and to provide data privacy. This approach, however, degrades the accuracy of the computations \cite{soleymani2021analog,soleymani2020privacy}. In fact, this is a limitation of many other problems such as verifiable computing and machine learning \cite{ghodsi2017safetynets,ali2020polynomial}. Recently, several works have extended LCC to the analog domain to address these challenges \cite{subramaniam2019collaborative,jahani2020berrut,soleymani2021analog,soleymani2020privacy}, but they either focus only on straggler-mitigation \cite{jahani2020berrut}, privacy \cite{soleymani2021analog,soleymani2020privacy} or Byzantine-robustness \cite{subramaniam2019collaborative}. An interesting open problem is to design a framework that jointly tackles these three challenges in the analog domain. 
\end{itemize}

\emph{Challenges and open problems in private federated learning:}
\begin{itemize}
\item Secure and Byzantine-robust aggregation in federated learning: While BREA \cite{so2020byzantine} has considered secure aggregation and mitigating Byzantine users jointly, it has only focused on the i.i.d.~setting. Extending BREA to the non-i.i.d.~setting is an interesting future direction. The main challenge in this direction is to determine whether the updates that may seem deviating are due to the users having non-i.i.d.~data or because of Byzantine users sending erroneous updates. 
\item Secure aggregation and multi-round secure aggregation: 
There are many open problems related to the multi-round secure aggregation problem introduced in \cite{so2021securing}. While the secure aggregation protocols are believed to protect the privacy the of the individual users, it is not clear whether such protocols ensure privacy in the information-theoretic sense. Specifically, the secure aggregation protocols ensure that the server only learns the aggregate model of the users. However, the aggregate model of the users may still reveal information about the individual users and characterizing such a leakage is an important problem. While Multi-RoundSecAgg provides a trade-off between between the multi-round privacy, the average aggregation cardinality, and the aggregation fairness, investigating the optimality of Multi-RoundSecAgg remains an open problem.  
\end{itemize}

\bibliographystyle{IEEEtran}
\bibliography{referencesPrivateComputing,CSPIR,PIRref,PIRj,FL-DP-Ref,PIR-papers-su}

\end{document}